# Combining optical and magnetic resonance spectroscopies to probe charge recombination via triplet excitons in organic solar cells

*Alberto Privitera[1,2#], Jeannine Grüne[3#], Akchheta Karki[4], William K. Myers[5], Vladimir Dyakonov[3], Thuc-Quyen Nguyen[4], Moritz K. Riede[1], Richard H. Friend[6], Andreas Sperlich[3*] and Alexander J. Gillett[6*].*

[1]Clarendon Laboratory, University of Oxford, Parks Road, Oxford, UK.

[2]Department of Chemistry, University of Torino, Via Giuria, Torino, Italy.

[3]Experimental Physics 6, Julius Maximilian University of Würzburg, Am Hubland, Würzburg, Germany.

[4]Centre for Polymers and Organic Solids, Department of Chemistry and Biochemistry, University of California at Santa Barbara, USA.

[5]Centre for Advanced ESR, Inorganic Chemistry Laboratory, University of Oxford, South Parks Road, Oxford, UK.

[6]Cavendish Laboratory, University of Cambridge, JJ Thomson Avenue, Cambridge, UK.

[#]These authors contributed equally.

[*]Corresponding authors: Alexander J. Gillett: E-mail: ajg216@cam.ac.uk; Andreas Sperlich: E-mail: sperlich@physik.uni-wuerzburg.de.

**Organic solar cells (OSCs) have recently shown a rapid improvement in their performance, bringing power conversion efficiencies (PCEs) closer to the point where commercial applications of the technology become viable. However, the low open-circuit voltage ($V_{OC}$) of OSCs relative to their optical gap still limits PCEs to below 20%. A key factor contributing to the large $V_{OC}$ deficit in OSCs is non-radiative recombination to spin-triplet excitons, which is widely, but not universally, observed in blends using both fullerene and non-fullerene electron acceptors. Here, we present an experimental framework that combines time resolved optical and magnetic resonance spectroscopies to detect triplet excitons and identify their formation mechanisms. We apply our methodology to two well-studied polymer:fullerene systems, PM6:$PC_{60}$BM and PTB7-Th:$PC_{60}$BM, enabling us to selectively investigate distinct triplet formation pathways. In contrast to the more efficient non-fullerene acceptor systems that show only triplet states formed *via* non-geminate recombination, the fullerene systems also show significant triplet formation *via* geminate processes. We associate this with electrons trapped at the isolated fullerenes that sit within the alkyl sidechains of the donor polymers. Thus, our model study demonstrates how these complex and overlapping processes can be successfully deconvoluted to reveal the intricacies of triplet generation dynamics in OSC blends.**

Driven by the recent development of non-fullerene electron acceptor (NFA) materials, the power conversion efficiencies (PCEs) of organic solar cells (OSCs) have rapidly increased, and now exceed 18%[1–5]. However, the photon energy loss in OSCs, defined as the difference between the optical gap ($E_g$) and the energy of the extracted charges ($qV_{OC}$)[6], remains high and is now the main factor limiting OSC performance[7,8]. The primary cause is excessive non-radiative recombination in OSCs[9], which impacts the open-circuit voltage ($V_{OC}$) of the solar cell by reducing the charge carrier lifetime from the intrinsic radiative limit[10–12]. This non radiative voltage loss ($\Delta V_{nr}$) can be directly calculated from the electroluminescence external quantum efficiency ($EQE_{EL}$) of the solar cell run at moderate forward bias as a light-emitting diode:

$$\Delta V_{nr} = \frac{-k_B T}{q} \ln(EQE_{EL}) \tag{1}$$

where $k_B$ is the Boltzmann constant, $T$ is temperature, and $q$ is the elementary charge. Here, the $EQE_{EL}$ can be further separated into the different contributions[13]:

$$EQE_{EL} = \gamma \Phi_{PL} \chi \eta_{out} \tag{2}$$

where $\gamma$ is the charge balance factor, $\Phi_{PL}$ is the photoluminescence quantum efficiency, $\chi$ is the fraction of recombination events that can decay radiatively (spin-singlet excitations) and $\eta_{out}$ is the photon out-coupling efficiency. To address the low $EQE_{EL}$ of OSCs, recent studies have focused on improving $\Phi_{PL}$ of spin-singlet excitations in the donor:acceptor blend[7,14–16]. However, the optimum optical gap for a single junction solar cell lies in the near-infrared spectral region[10,17] where most OSCs show stronger multi-phonon non-radiative recombination rates, termed the energy gap law[9,18]. As such, this may limit the scope for increasing $\Phi_{PL}$ of spin-singlet states in such systems[19].

Recently, the recombination of charge carriers *via* spin-triplet excitons has been identified as another significant non-radiative voltage loss pathway in both fullerene and NFA OSCs[20–23]. For example, in the benchmark PM6:Y6 system[24], the fraction of charge carriers that recombine *via* the triplet exciton ($T_1$) of the low $E_g$ component, Y6, is ~90%[20]; comparable $T_1$ recombination fractions have also been reported in fullerene acceptor OSCs[25,26]. As a result, $\chi$ in Equation 2 is limited to 0.1 and the $EQE_{EL}$ of the PM6:Y6 blend is reduced by a factor of 10, lowering the $V_{OC}$ by ~60 mV. We note that to achieve a $V_{OC}$ gain comparable to eliminating recombination via $T_1$, $\Phi_{PL}$ of the low $E_g$ component, which is considered to provide the limit for $\Phi_{PL}$ of the blend when recombination can proceed via the lowest energy singlet exciton ($S_1$)[7,15], should be raised by a factor of 10. To explore the feasibility of such an improvement in luminescence efficiency, we have measured $\Phi_{PL}$ for a neat film of the widely used NFA Y6. Here, we obtain $\Phi_{PL}$ =2%, which would necessitate an increase in $\Phi_{PL}$ to ~20%. A $\Phi_{PL}$ of this magnitude would be unprecedented among fluorescent organic small molecules with an emission peak around 950 nm[19], moreover in a molecule that can also operate as an electron acceptor in a highly efficient OSC. Thus, whilst enhancing $\Phi_{PL}$ can provide incremental improvements in device performance, it is unlikely to alone yield the step-change in $V_{OC}$ required for PCEs of >20% to be realised. Furthermore, triplet states have also been implicated in the degradation of both fullerene and NFA OSC blends, potentially presenting a fundamental barrier to commercial applications which require photovoltaic modules with long term stability[27–30]. Therefore, eliminating recombination *via* $T_1$ should now be a key focus for further improving the $V_{OC}$ and operational lifetimes of OSCs. However, we note this remains an understudied topic in the field[23,31–33]. This is likely due to the difficulty in detecting and characterising spin-triplet states in organic semiconductors, as they are generally optically dark and, in the context of OSCs, often short lived due to the presence of rapid annihilation processes[20].

We present here an experimental framework for probing triplet excitons in OSCs to assist with the task of engineering out recombination *via* $T_1$. We consider that there are three main ways in which $T_1$ can be created in OSCs: (1) direct intersystem crossing (ISC) from un-dissociated $S_1$ states (Figure 1a); (2) back charge transfer (BCT) from geminate spin-triplet charge transfer ($^3$CT) states (Figure 1b); (3) BCT from $^3$CT states formed *via* non-geminate recombination (Figure 1c). In general, any $T_1$ states formed will ultimately relax to the lowest energy $T_1$ in the system. But, if molecular $T_1$ states are energetically higher than the $^3$CT states, $T_1$ will not be formed[34]. As $^3$CT states are readily converted to spin-singlet CT states ($^1$CT) through the hyperfine interaction (HFI)[35], their presence is not expected to significantly impact device performance[20]. Therefore, $T_1$ states become a problem if they are energetically below the CT states. However, as the offset between CT states and the lowest energy $S_1$ is typically small (<0.2 eV) to reduce energy losses associated with charge generation[7,9,14], the molecular $T_1$ states almost always lie below the CT states due to the large $S_1$-$T_1$ energy gap in most organic semiconductors with localised molecular excitons[36–38]. Furthermore, as most OSC blends comprise at least one donor and one acceptor component, with ternary systems containing an additional donor or acceptor[2–5,39,40], there is the potential for $T_1$ states to be formed on any of these materials through each of the three pathways presented above. Thus, fully understanding these complex and overlapping mechanisms will require the application of multiple experimental techniques, each targeting a specific subset of the possible $T_1$ formation pathways.

To achieve this, we propose the combination of optical and magnetic resonance spectroscopies, which have previously been utilised for investigating $T_1$ states in organic semiconductors[20,25,41–43]. Specifically, we present a framework to investigate recombination *via* $T_1$ through three complementary methods: transient absorption (TA), time-resolved electron paramagnetic resonance (trEPR), and photoluminescence detected magnetic resonance (PLDMR) spectroscopies. In Table 1, we present a summary of the three different techniques, as well as the $T_1$ recombination pathways that they can detect; a more detailed discussion on each of these techniques is provided with the corresponding experimental results (*vide infra*).

To develop our experimental framework, we have chosen to examine two model fullerene OSC systems: PM6:$PC_{60}$BM and PTB7-Th:$PC_{60}$BM (chemical structures in Figure 1d). These blends give PCEs of 7.4% and 7.5%, respectively; further information is given in Figure S1. The polymers PM6 and PTB7-Th have been chosen as they are commonly used donor materials in efficient fullerene and NFA OSCs[3,24,39,40,44–47]. We have opted to use fullerene acceptors for two key reasons. First, unlike NFAs which exhibit strong spectral features in TA[20,48,49], the fullerene component does not make any significant contribution to the observed TA spectrum in the visible and near infrared probe regions. Thus, the use of fullerene blends avoids the complex superposition of the polymer and NFA spectral features and dynamics, simplifying the data interpretation. Second, in many NFA blends, the geminate BCT pathway to $T_1$ states is not observed in trEPR[20]. Conversely, fullerene blends often show geminate BCT $T_1$ formation[23,42,50,51]. Thus, fullerene acceptor blends are the ideal model systems to demonstrate how it is possible to probe the three main $T_1$ formation mechanisms, clarifying the strength of our approach and the complementarity of optical and magnetic resonance techniques. For brevity, we focus here on PM6:$PC_{60}$BM, but a full discussion of the results for PTB7-Th:$PC_{60}$BM is presented in the SI.

We begin by using TA to explore $T_1$ formation. TA has been widely used to explore the photophysical processes occurring in OSCs[20,49,52–55], as it is able to provide insights into the evolution of both optically bright and dark states on timescales spanning femtoseconds to milliseconds. Thus, TA is well suited to probing optically dark $T_1$ states in OSCs as the distinct $T_1$ photo-induced absorption (PIA) signatures, typically located in the near-infrared (NIR) spectral region[20–22,25,26,56], provide a clear fingerprint for the presence and molecular location of these states. Furthermore, TA can, in theory, distinguish between the presence of a monomolecular (direct ISC or geminate BCT) or bimolecular (non-geminate BCT) $T_1$ formation pathways through the fluence dependence of $T_1$ generation; monomolecular pathways show no fluence dependence[56], whilst bimolecular events exhibit a strong fluence dependence[20,25,26]. However, in the case where significant bimolecular pathways are present, the fluence dependent behaviour of $T_1$ formation will dominate, masking any underlying monomolecular processes. Thus, as many fullerene and NFA OSCs blends demonstrate non-geminate BCT $T_1$ formation[20–22,25,26], TA can, in general, only be reliably used to detect the non-geminate pathway.

In Figure 2a, we present the TA from the NIR region of PM6:$PC_{60}$BM. At 0.2-0.3 ps after photoexcitation at 600 nm, we observe the presence of a PIA centred around 1175 nm. Through comparison to the TA of a neat PM6 film (Figure S2), we attribute this feature to the $S_1$ state of PM6. The PM6 $S_1$ is rapidly quenched within a picosecond, indicating ultrafast electron transfer to $PC_{60}$BM. Subsequently, over timescales of hundreds of picoseconds, we notice the formation of a new PIA band peaking at the edge of our probe range around 1650 nm. As the $T_1$ PIA for $PC_{60}$BM has previously been reported at 720 nm[57,58], we attribute this new PIA to $T_1$ located on PM6. We also observe a strong fluence dependence for the formation of this new PIA (Figure 2b), indicating that the PM6 $T_1$ states observed are generated through the non-geminate BCT process. We make similar observations in the TA of PTB7-Th:$PC_{60}$BM (Figure S4), confirming that non-geminate BCT $T_1$ formation is also present in this system. In the highest fluence measurement for PM6:$PC_{60}$BM presented here (6 μJ $cm^{-2}$), we note that the $T_1$ PIA intensity peaks around 300 ps, before decaying again; this can be attributed to triplet-charge annihilation (TCA). As the rate of TCA depends on the charge carrier density in the blend film (as well as the charge carrier mobility[25]), it is expected to become more prominent on sub-nanosecond timescales under higher excitation fluences[20,26,59]. Indeed, TCA is the primary non-radiative quenching pathway of $T_1$ in OSCs and is therefore directly responsible for the increased non-radiative voltage losses in OSCs with significant $T_1$ formation[20,23,37]. Incidentally, the observation of substantial charge carrier recombination to $T_1$ on sub-nanosecond timescales, even at moderate fluences of a few μJ $cm^{-2}$, underlines the importance of performing TA measurements at very low excitation fluences (<1 μJ $cm^{-2}$), if reliable data that accurately represents the photophysics of operational OSCs under 1 Sun illumination intensities is to be obtained.

Having explored the non-geminate BCT pathway, we now investigate the geminate BCT and direct ISC pathways for $T_1$ formation. To achieve this, we turn to trEPR spectroscopy. trEPR typically has a time resolution on the order of hundreds of nanoseconds and is sensitive to the presence of states with unpaired spins[60,61]; in OSCs, this primarily includes $T_1$ states, spin-correlated radical pairs (which can be considered analogous to CT states in an OSC blend), and free charge carriers[42,50,51]. With a focus on $T_1$ states, trEPR not only provides information on the molecular location of $T_1$ and the local structure in the blend through the zero-field splitting (ZFS) parameters of the spin Hamiltonian, but also on the $T_1$ formation mechanism through the spin-polarisation of the signal[60,62]. In trEPR spectroscopy, the spin-polarisation results from non-Boltzmann population of the triplet sublevels, which manifests as a characteristic polarisation pattern of absorptive (*a*) and emissive (*e*) microwave-induced EPR transitions between the three triplet sublevels. For example, direct ISC mediated by spin-orbit coupling (SOC) from $S_1$ results in a spin selective population of the zero-field triplet sublevels (Table S1)[63], which in turn is converted into a polarised population of the three high-field triplet sublevels $T_+$, $T_0$ and $T_-$. By applying microwave irradiation, transitions between these sublevels result in an *aaaeee, eeeaaa*, *eeaeaa*, *aaeaee, aeaeae or eaeaea* polarisation pattern. In contrast, $T_1$ created through the geminate BCT mechanism will possess a characteristic *aeeaae* or *eaaeea* signature resulting from ISC primarily mediated by the HFI with paramagnetic nuclei (mainly protons)[27,64]. In contrast, non-geminate recombination does not produce spin polarisation as the spin-statistical recombination of uncorrelated free charge carriers to $T_1$ *via* $^3$CT results in an equal population of the $T_+$, $T_0$ and $T_-$ sublevels. The equal sublevel population will establish Boltzmann population within the spin-lattice relaxation time, but in the low-field (~330 mT) regime that we explore here with X-band EPR, the spin polarisation is too low to be detected with trEPR[62]. Therefore, the non-geminate BCT pathway does not induce a sufficiently high spin-polarisation in $T_1$, and states populated through this mechanism are not observed in trEPR. Thus, trEPR provides an excellent complement to TA spectroscopy, in which only the non-geminate pathway can be reliably detected.

In Figures 3a-3b, we show the trEPR spectra and associated simulations of the PM6:$PC_{60}$BM film taken at two representative time points (1 and 5 μs) after excitation at 532 nm. A summary of the best fit simulation parameters for the blends studied is included in Table 2, with more detailed information on all samples in Table S1. At 1 μs, we observe an intense and spectrally narrow *aeae* feature centred at ~346 mT, assigned to CT states; by 5 μs, this evolves into a pure *a* signal, indicative of free charges[61]. Conversely, the broader signal between 290-410 mT is assigned to $T_1$ states. From the best-fit spectral simulation at 1 μs, an *eeeaaa* polarisation pattern is obtained, confirming that the $T_1$ states are formed via SOC-ISC; this results from photogenerated $S_1$ states that do not undergo charge transfer at the donor:acceptor interface. The $T_1$ ZFS parameters, $D$ = 1300 MHz and $E$ = 140 MHz, are comparable to the those obtained at 1 μs in a neat PM6 film (Figure S5) and are significantly larger than those found for a neat $PC_{60}$BM film (Figure S6; $D$ = -237 MHz and $E$ = 39 MHz).

Therefore, we assign this feature to the $T_1$ of PM6. In contrast, by 5 μs, there is a clear evolution of the polarisation pattern, and a more complex spectrum is observed. We attempt to simulate this new spectrum using a single SOC-ISC component but find through examining the residual that this species alone is not sufficient to obtain a high-quality fit (Figure S7). An excellent fit is only obtained when two $T_1$ species ($D$ = 1220 MHz and $E$ = 40 MHz) are included in the simulation with distinct *aeaeae* and *eaaeea* polarisation patterns. The *aeaeae* species is the same as the PM6 $T_1$ formed *via* SOC-ISC at 1 μs, with the apparent spectral inversion attributed to an unequal rate of decay from the three high-field triplet sublevels[65]. However, the new *eaaeea* contribution represents $T_1$ formed *via* the geminate BCT mechanism, confirming that this pathway is also present in the PM6:$PC_{60}$BM blend. Interestingly, we observe the presence of geminate BCT $T_1$ states at the earlier time of 1 μs in PTB7-Th:$PC_{60}$BM (Figure S9), compared to 5 μs in PM6:$PC_{60}$BM. The slower geminate BCT $T_1$ formation in PM6:$PC_{60}$BM is also correlated with the faster evolution of CT states into free charges in this system at 80 K; we observe only CT states, not free charges, in PTB7-Th:$PC_{60}$BM at both 1 and 5 μs. This is consistent with the idea that the slower separation of photo-generated $^1$CT states into charges provides more opportunity for subsequent HFI-induced spin-mixing with the $^3$CT state, followed by terminal BCT to a molecular $T_1$ state[20].

We note that the presence of triplet excitons generated by geminate BCT in fullerene acceptor blends, as reported here by us and others[23,42,50,51], is in clear contrast to the results obtained for the NFA blends[20]. These observations can be rationalized by the high solubility of isolated fullerenes within the alkyl side chains of the donor polymer, resulting in the formation of mixed polymer/fullerene regions[66–70]. Efficient solar cell operation is obtained when there is excess fullerene that forms local 5-10 nm fullerene inclusions, requiring electron transfer from mixed fullerene/sidechain regions to the pure fullerene regions. If the acceptor concentration in these mixed regions is below the percolation threshold for efficient electron transport, charge separation will be impeded and geminate recombination will result[71,72]. Thus, we propose that poorly-connected fullerenes provide the opportunity for geminate $^3$CT formation *via* HFI-ISC from $^1$CT states on nanosecond timescales[56,73], followed by BCT to $T_1$, increasing losses *via* $T_1$ states. In contrast, many efficient NFA OSCs have been shown to possess good phase purity[74–77], which has previously been shown to facilitate CT state dissociation and reduce BCT $T_1$ formation[78]. Thus, it appears that engineering good phase purity in the donor:acceptor bulk heterojunction could be helpful for engineering out BCT $T_1$ generation pathways in OSCs.

In contrast to TA and trEPR, PLDMR spectroscopy is generally employed as a steady-state technique[79], though transient iterations with a time resolution on the order of tens of nanoseconds are available[80]. Continuous illumination can also yield spin polarisation of the triplet sublevels by unequal recombination rates or triplet accumulation, with subsequent triplet-triplet annihilation processes. Furthermore, whilst trEPR directly detects reflection changes in the applied microwaves following laser excitation, PLDMR uses optical detection[30,43,79,81]. Therefore, the experimental sensitivity is greatly enhanced and PLDMR can detect any triplet excitations that are coupled to the photoluminescence of the sample (for example, *via* the triplet-triplet annihilation of two molecular $T_1$ which reforms, among other possibilities, one bright $S_1$ state and one dark spin-singlet ground state)[79], not just those that are highly spin polarised. In addition, the enhanced sensitivity of PLDMR allows for 'half-field' (HF) signals to be readily resolved; HF transitions represent a first-order forbidden $\Delta m_S = \pm 2$ transition between the $T_+$ and $T_-$ sublevels that is enabled by the strong dipolar interaction between the two localised electron spins of a molecular $T_1$ state[30,43]. These signals provide an additional tool for determining the molecular location of a $T_1$ state, since their magnetic field position depends on the strength of the dipolar interaction[82]. Therefore, HF signals are particularly useful when working with systems where the ZFS parameters of the donor and acceptor $T_1$ states are similar, which is often the case in NFA OSC blends[20].

In Figure 4 (enlarged individual spectra are presented in Figures S12-S16), we display the PLDMR spectra of neat films of $PC_{60}BM$ (red), PM6 (light blue) and the PM6:$PC_{60}BM$ blend (dark blue). A summary of the simulation parameters used for the films is shown in Table 2, with the full spectral simulations and ZFS parameters for each sample provided in Figure S17 and Table S2. The spectra consist of a full-field (FF) region (260-410 mT), corresponding to $\Delta m_S = \pm 1$ transitions, and $\Delta m_S = \pm 2$ HF signals (160-172 mT). Beginning with neat $PC_{60}BM$ (Figure 4, red), we observe a relatively narrow $T_1$ feature in the FF spectrum between 320-350 mT, which can be described with the ZFS parameters $D$ = 360 MHz and $E$ = 50 MHz. Two additional features (sharp negative signals) are superimposed on the $T_1$ signal at 336.25 mT (g = 2.0012) and 336.65 mT (g = 2.0040), seen more clearly in Figure S9c. We assign these to the anion $PC_{60}BM^-$, as already known from literature, and the cation $PC_{60}BM^+$, respectively[83–85]. In addition, a HF signal due to $PC_{60}BM$ $T_1$ states is also detected at 168.1 mT, which can be easily distinguished from the HF signals of the polymer in the blends.

For PM6 (light blue), we observe a broad $T_1$ spectrum between 280-390 mT, corresponding to ZFS parameters $D$ = 1500 MHz and $E$ = 70 MHz. In contrast to trEPR, this spectrum shows a considerable ordering factor, $\lambda$, which provides information on the preferred orientational distribution of the molecules in the sample and is reflected by the 'wings' in the PLDMR spectrum (Figure S18, Table S2)[62,86,87]. When blending PM6 with $PC_{60}BM$, the broader polymer triplet is again clearly visible in the FF and HF signals. The PM6 $T_1$ ZFS parameters remain the same before and after blending with $PC_{60}BM$ and the ordering factors change only slightly (Table S2). This is in clear contrast to PTB7-Th:$PC_{60}BM$ (Figure S11), where the ZFS parameters and ordering factors of PTB7-Th change significantly upon blending. Thus, the ordering of the polymer chains in PM6 is less disrupted upon mixing with $PC_{60}BM$, when compared to PTB7-Th (see SI for more details). Additionally, both blends show a positive ($\Delta PL/PL$ = 0) CT state peak with two negative signals at g = 2.0012 and g = 2.0037 in PTB7-Th:$PC_{60}BM$, and g = 2.0012 and g = 2.0034 in PM6:$PC_{60}BM$. The lower g value is identical to the negative polaron $PC_{60}BM^{-}$, as detected in neat $PC_{60}BM$ (Figure S12c), whilst the higher g value likely represents positive polarons on the respective polymer[88].

In contrast to trEPR, the additional experimental sensitivity of PLDMR enables us to resolve the weaker $PC_{60}BM$ $T_1$ signal in the HF (168.1 mT) regions in both blends (Figures S12a, S13a); only the more intense $T_1$ features of the donor polymers are seen in the corresponding trEPR data (Figures 3 and S9). Thus, without PLDMR, minority $T_1$ generation pathways could easily be missed. When considering the formation mechanism for $PC_{60}BM$ $T_1$ states in the blends, we note that they are generally too high in energy ($T_1$ ~1.5 eV) to be populated by BCT from the CT states in low band gap polymer:fullerene systems with large $S_1$-CT energetic offsets (as is the case in the blends studied here)[34]. Furthermore, as the $PC_{60}BM$ $T_1$ will be higher in energy than the $T_1$ states of PM6 and PTB7-Th, assuming a typical $S_1$-$T_1$ energy gap of ~0.6-1 eV in most conjugated polymers[38], any $PC_{60}BM$ $T_1$ states formed near the donor:acceptor interface would be expected to relax into the lower lying polymer $T_1$ state. Therefore, we conclude that the $PC_{60}BM$ $T_1$ states observed must be located in isolated $PC_{60}BM$ domains, which are often found in polymer:fullerene bulk heterojunction blends[89,90]. This observation reinforces the importance of ensuring that domain sizes are on the order of the exciton diffusion length to enable efficient charge generation and suppressed $T_1$ formation *via* direct SOC-ISC from undissociated $S_1$ states[91].

In this work, we have shown that through leveraging the strengths of TA, trEPR, and PLDMR spectroscopies, a complete picture of $T_1$ generation pathways in OSCs can be obtained. This strategy is fully applicable to both fullerene and NFA blends and will prove valuable in the task of engineering out of recombination *via* $T_1$ states in OSCs. Through applying this methodology to two model OSC blends, we have demonstrated that it is possible to unravel the intricacies of spin-triplet physics in OSCs by identifying both the molecular localisation and generation mechanism for the wide range of $T_1$ states found in these systems. Consequently, we have shown that fullerene blends frequently exhibit the geminate BCT $T_1$ formation mechanism[23,42,50,51]. This is in clear contrast to the more efficient NFA OSC systems where this pathway appears to be absent, in agreement with their improved performance[20]. We propose that the geminate BCT mechanism is associated with isolated fullerene molecules trapped in alkyl side chains of the donor polymers[66–70], suggesting that engineering good purity in the donor and acceptor phases is key for supressing this process. Thus, we anticipate that this framework will also be particularly useful for analysing $T_1$ loss mechanisms in ternary systems comprised of both fullerene and NFAs, which have demonstrated some of the highest PCEs to date[39,40].

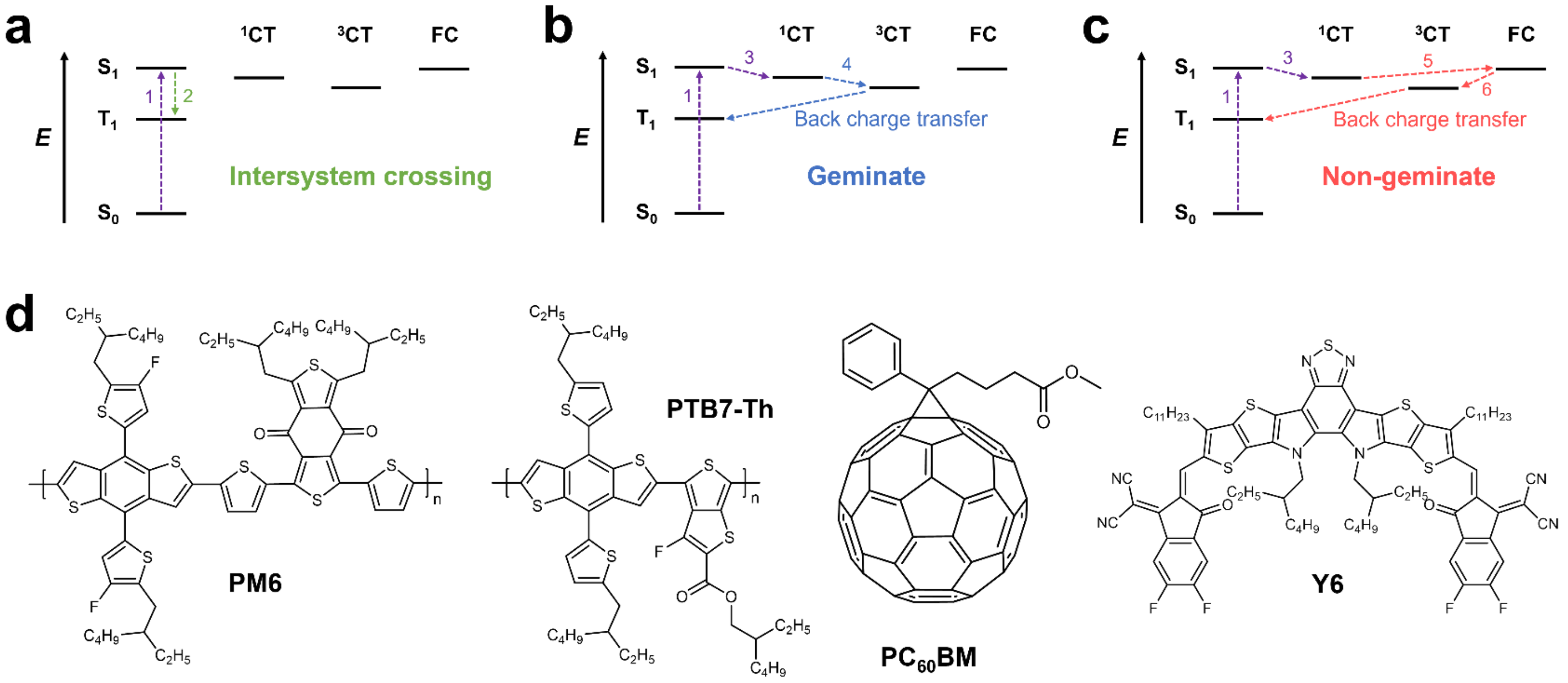

**Figure 1: Triplet formation pathways and organic solar cell materials studied. (a)** Schematic of the direct intersystem crossing pathway for $T_1$ formation in organic solar cells. After optical excitation into $S_1$ (1), charge transfer to the $^1$CT state does not occur and $T_1$ is formed by intersystem crossing instead (2). **(b)** Schematic of the geminate BCT pathway for $T_1$ formation in organic solar cells. After optical excitation into $S_1$ (1), charge transfer to the $^1$CT state successfully occurs (3). However, the $^1$CT state does not separate into free charges (FC) and instead undergoes spin-mixing to form $^3$CT states (4). These $^3$CT states then undergo a spin-allowed back charge transfer process to form molecular $T_1$ states. **(c)** Schematic of the non-geminate BCT pathway for $T_1$ formation in organic solar cells. After optical excitation into $S_1$ (1), charge transfer to the $^1$CT state successfully occurs (3). $^1$CT then separates into free charges (5). Spin-statistical non-geminate recombination of free charges leads to the formation of $^3$CT states, which can then undergo back charge transfer to form molecular $T_1$ states. **(d)** The chemical structures of the material used in this study.

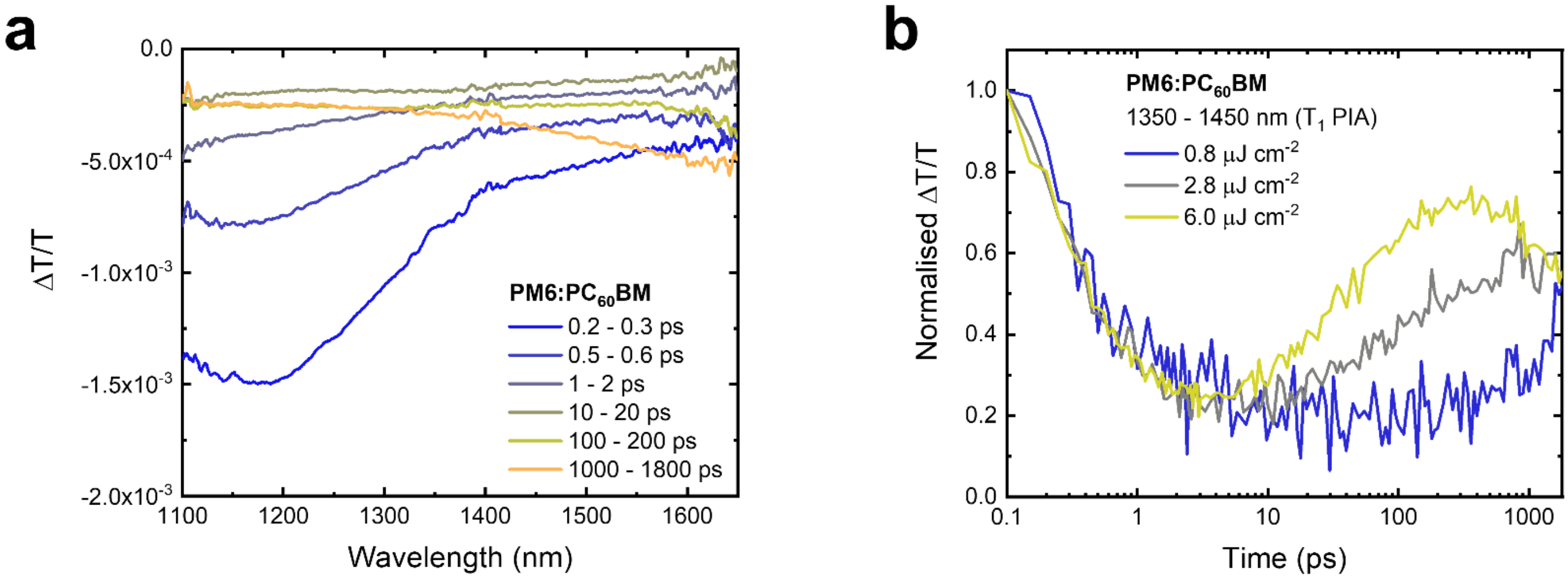

**Figure 2: Transient absorption spectroscopy of the PM6:$PC_{60}$BM blend. (a)** The TA spectra of a PM6:$PC_{60}$BM blend film, excited at 600 nm with a fluence of 2.8 μJ $cm^{-2}$. The PM6 $S_1$ PIA centred at 1175 nm decays within the first picosecond due to electron transfer to $PC_{60}$BM. Over hundreds of picoseconds, a new PIA band around 1650 nm begins to grow in, indicating recombination into PM6 $T_1$ states. **(b)** The TA kinetics a PM6:$PC_{60}$BM blend film, excited at 600 nm with varying fluence. The fluence dependence of the $T_1$ PIA growth shows that $T_1$ formation occurs following the bimolecular recombination of free charge carriers. All TA measurements were performed at 293 K.

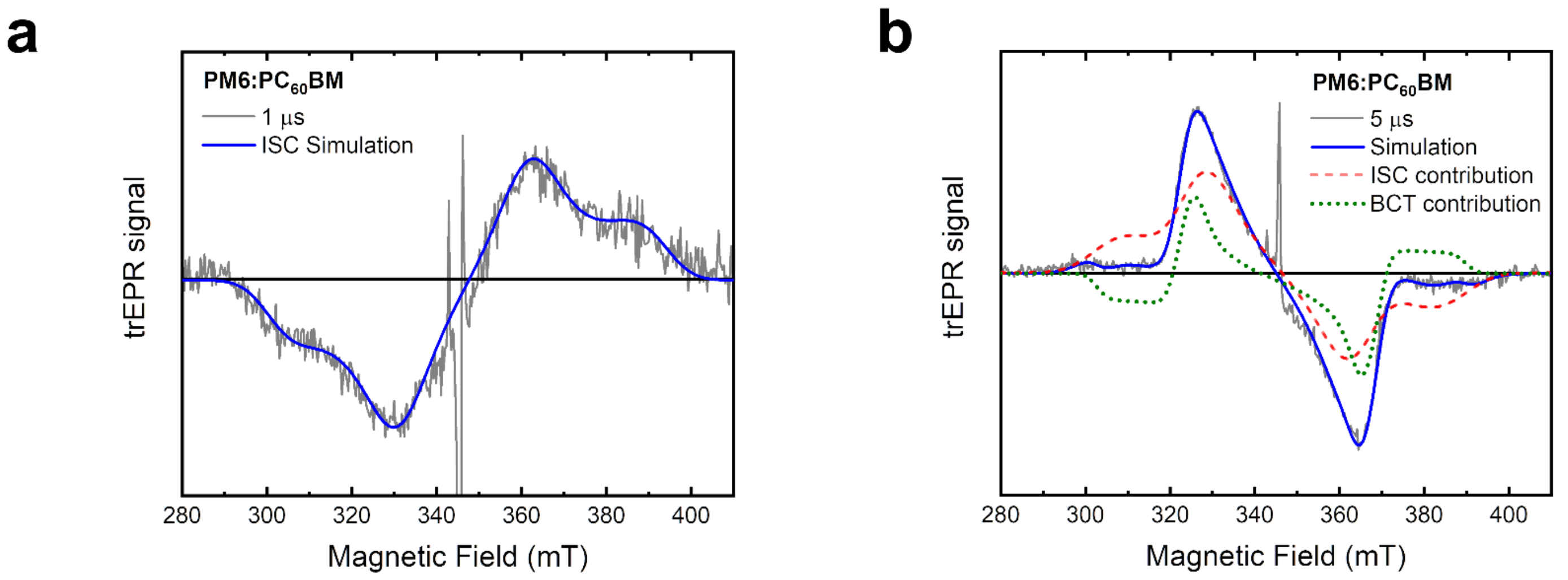

**Figure 3: Time resolved EPR spectra of the PM6:$PC_{60}$BM blend. (a)** The trEPR spectrum of a PM6:$PC_{60}$BM blend film at 1 μs, excited at 532 nm. Absorption (*a*) is up, emission (*e*) is down. The central *aeae* species is assigned to a CT state. The PM6 $T_1$ species can be simulated with a single *eeeaaa* component, indicating it is formed *via* direct ISC from undissociated $S_1$ states. **(b)** The trEPR spectrum of a PM6:$PC_{60}$BM blend film at 5 μs, excited at 532 nm. The *aeae* CT state has now evolved into a single *a* signal, indicative of free charges. The PM6 $T_1$ species requires the use of two $T_1$ contributions to successfully describe the observed spectrum. The first is an *aeaeae* component, which is the same ISC $T_1$ state as the 1 μs spectrum, except inverted. The second is an *eaaeea* component, which is characteristic of $T_1$ states formed *via* the geminate BCT mechanism. All trEPR measurements were performed at 80 K.

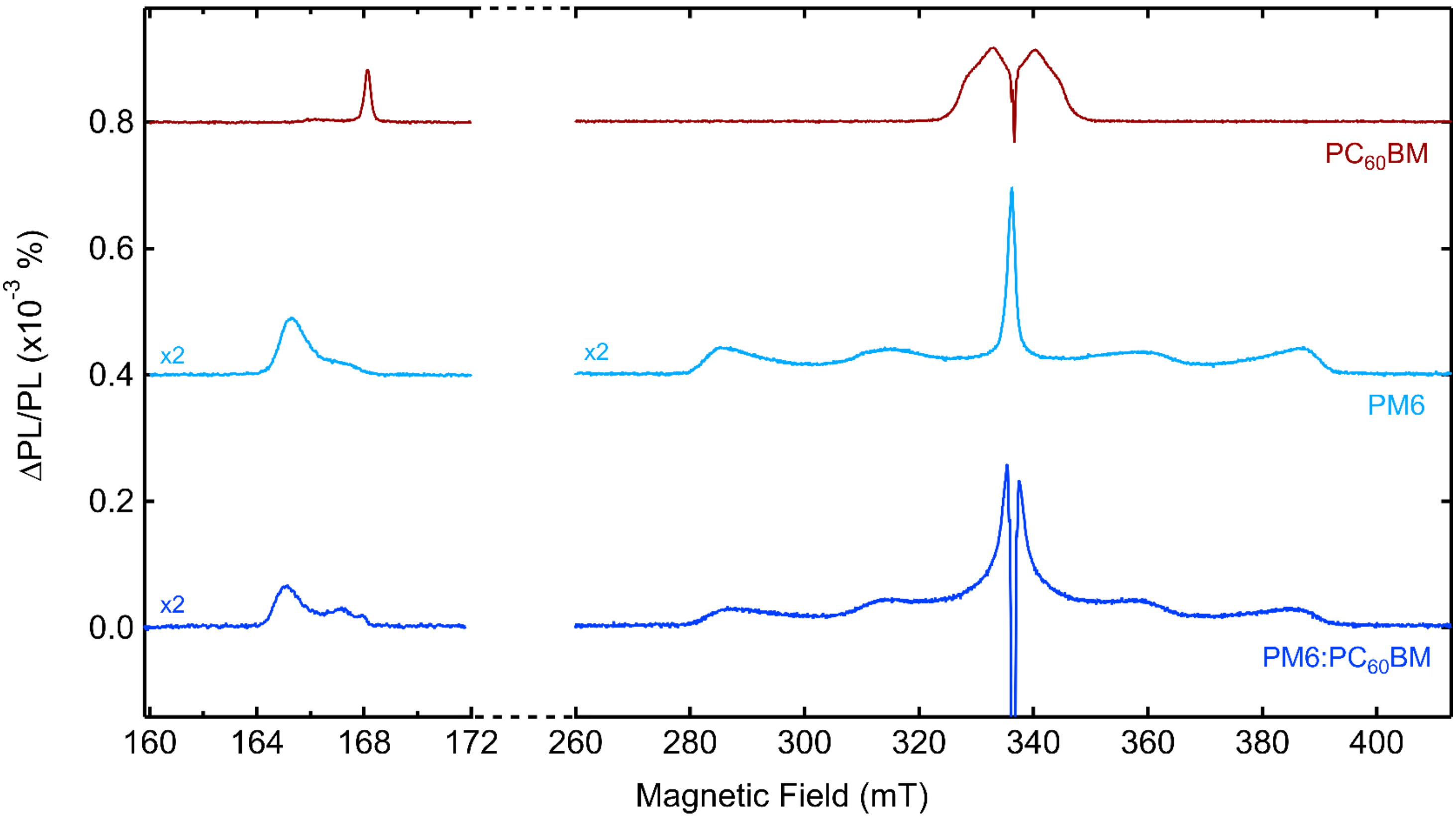

**Figure 4: Photoluminescence-detected magnetic resonance spectra of PM6, $PC_{60}BM$, and the corresponding blend film. (Red)** The half-field (160-172 mT) and full-field (260-410 mT) PLDMR spectrum of a neat $PC_{60}BM$ film. The HF $T_1$ signal of $PC_{60}BM$ is visible at 168.1 mT, whilst the FF $T_1$ and polaron features are present between 320-350 mT. **(Light blue)** The HF and FF PLDMR spectrum of a neat PM6 film. The HF $T_1$ signal of PM6 is visible at 165.3 mT, whilst the FF $T_1$ is the broad feature spanning 280-390 mT. **(Dark blue)** The HF and FF PLDMR spectrum of the PM6:$PC_{60}BM$ blend film. The HF $T_1$ signals of PM6 and $PC_{60}BM$ are visible at 165.3 mT and 168.1 mT, respectively. The FF PM6 $T_1$ is visible between 280-390 mT. All PLDMR spectra were acquired at 10 K.

| Technique | $T_1$ mechanisms detected | Comments |
|---|---|---|
| Transient absorption | – Non-geminate BCT<br>– Geminate BCT<br>– Direct ISC from $S_1$ | – Can easily assign non-geminate BCT $T_1$ through fluence dependence measurements<br>– Cannot easily distinguish between geminate BCT and direct ISC $T_1$ as both are fluence-independent processes<br>– Donor and acceptor $T_1$ PIAs can be differentiated through triplet sensitisation experiments<br>– Can probe timescales from fs to ms<br>– Generally performed at room temperature<br>– Excitation fluences can be comparable to 1 Sun<br>– Can be quantitative through knowledge of the absorption cross section for $T_1$ |
| Transient electron paramagnetic resonance | – Geminate BCT<br>– Direct ISC from $S_1$ | – $T_1$ spin polarisation allows simple differentiation between population mechanisms<br>– Donor and acceptor $T_1$ can be easily differentiated if ZFS *D*-parameter is significantly different<br>– Can probe timescales from hundreds of ns to μs<br>– Very high excitation fluences (mJ pulse energies typically used)<br>– Good sensitivity, but only to spin-polarised species<br>– Performed at cryogenic temperatures (<80 K) |
| Photoluminescence-detected magnetic resonance | – Non-geminate BCT<br>– Geminate BCT<br>– Direct ISC from $S_1$ | –Very high sensitivity to all triplet states that can couple to the photoluminescence of the sample<br>– High sensitivity of optical detection easily allows differentiation of donor and acceptor $T_1$ states through ZFS *D*-parameters and half-field signals |
| | | – Non-trivial to differentiate between $T_1$ formation mechanism because it is employed as a steady-state technique<br>– Excitation power can be comparable to 1 Sun<br>– Often performed at cryogenic temperatures, but can also be used at room temperature |

**Table 1: A summary of the three techniques explored in this work.** This table identifies which $T_1$ formation mechanisms each technique can detect and provides other information relevant to their application in studying organic solar cell blends.

| Blend | Non-geminate BCT | Geminate BCT | SOC-ISC | *D, E* [MHz] | $T_1$ location |
|---|---|---|---|---|---|
| PM6:$PC_{60}$BM | TA | -<br>trEPR (5µs) | trEPR (1 µs)<br>trEPR (5µs) | 1300, 140<br>1220, 40 | D |
| | PLDMR | | | 1500, 70 | D, A (weak) |
| PTB7-Th:$PC_{60}$BM | TA | trEPR (1µs)<br>trEPR (5 µs) | trEPR (1 µs)<br>trEPR (5 µs) | 1050, 200<br>1143, 164 | D |
| | PLDMR | | | 1470, 180 | D, A (weak) |

**Table 2: A summary of the $T_1$ formation pathways detected by each method in the PM6:$PC_{60}$BM and PTB7-Th:$PC_{60}$BM blends**. All three blends exhibit the non-geminate BCT, geminate BCT, and SOC-ISC, $T_1$ formation pathways. TA and trEPR identify $T_1$ states on the donor (D) polymer, whilst PLDMR also detects a weak signature of $T_1$ states on the acceptor (A) fullerene material, likely formed *via* direct SOC-ISC from undissociated $PC_{60}$BM $S_1$ states.

# Methods

## OSC device fabrication

Indium tin oxide (ITO) patterned glass substrates were cleaned by scrubbing with soapy water, followed by sonication in soapy water, deionized (DI) water, acetone, and isopropanol for 20 minutes each. The substrates were dried using compressed nitrogen and placed in an oven overnight at 100 °C. The conventional architecture devices were made by treating the ITO substrates with UV-ozone for 15 minutes and spin-coating a layer of poly(3,4-ethylenedioxythiophene):poly(styrenesulfonate) (PEDOT:PSS, Clevios P VP Al 8043) at 3000 rpm for 40 s onto the ITO substrates in air. The substrates were then annealed in air at 150 °C for 20 minutes. Active layers were spin coated on top of the PEDOT:PSS layer inside a nitrogen filled glovebox following the recipes from previous reports[44,74,92]. The substrates were then pumped down under vacuum (<$10^{-7}$ torr), and a 5 nm thick Ca interlayer followed by a 100 nm thick Al electrode were deposited on top of the active layer by thermal evaporation using the Angstrom Engineering Series EQ Thermal Evaporator. In the case of inverted architecture devices, ZnO was used as the bottom transparent electrode (replacing PEDOT:PSS), where the ZnO solution was prepared in a nitrogen glovebox by mixing tetrahydrofuran and diethylzinc (2:1). The fresh ZnO solution was then spin-coated atop the clean ITO substrates at 4000 rpm for 30 seconds and then placed on a hotplate at 110 $^{0}$C for 15 minutes. Following active layer spin-coating, the inverted devices were pumped down under vacuum (<$10^{-7}$ torr), and 7 nm of $MoO_x$ and 100 nm thick Ag electrode were deposited on top of the active layer by thermal evaporation. The electrode overlap area was 0.22 $cm^2$ for both conventional and inverted devices. The active area of the device was determined using an optical microscope.

**OSC device testing**

Photovoltaic characteristic measurements were carried out inside a $N_2$ filled glove box. Solar-cell device properties were measured under illumination by a simulated 100 mW cm$^{-2}$ AM1.5 G light source using a 300 W Xe arc lamp with an AM 1.5 global filter. The irradiance was adjusted to 1 sun with a standard silicon photovoltaic cell calibrated by the National Renewable Energy Laboratory. No spectral mismatch correction was applied. A Keithley 2635A source measurement unit was used to scan the voltage applied to the solar cell between -2 to 1 V at a speed of 0.43 V/s with a dwell time of 46 ms. Scans were performed in both the forward and reverse directions, with no unusual behaviour observed. Between eight and 30 individual solar cell devices were tested for each blend reported. The error associated with the reported PCE values is ±0.2%.

**Photoluminescence quantum efficiency measurements**

The PLQE was determined using method previously described by De Mello *et al.*[93]. Samples were placed in an integrating sphere and photoexcited using a 658 nm continuous-wave laser. The laser and emission signals were measured and quantified using calibrated Andor iDus DU420A BVF Si and Andor CCD-1430 InGaAs detectors.

**TA spectroscopy**

TA was performed on a setup powered using a commercially available Ti:sapphire amplifier (Spectra Physics Solstice Ace). The amplifier operates at 1 kHz and generates 100 fs pulses centred at 800 nm with an output of 7 W. A TOPAS optical parametric amplifier (OPA) was used to provide the tuneable ~100 fs pump pulses. The probe was provided by a broadband IR non-collinear optical parametric amplifier (NOPA). The probe pulses are collected with an InGaAs dual-line array detector (Hamamatsu G11608-512DA), driven and read out by a custom-built board from Stresing Entwicklungsbüro. The probe beam was split into two identical beams by a 50/50 beamsplitter. This allowed for the use of a second reference beam which also passes through the sample but does not interact with the pump. The role of the reference was to correct for any shot-to-shot fluctuations in the probe that would otherwise greatly increase the structured noise in our experiments. Through this arrangement, very small signals with a $\frac{\Delta T}{T}$ =1x10$^{-5}$ could be measured.

**trEPR spectroscopy**

EPR samples were fabricated by spin-coating solutions under identical conditions to the optimised devices onto Mylar substrates, which were subsequently cut into strips with a width of 3 mm. To ensure the flexible Mylar substrates did not bend during the spin coating process, they were mounted onto rigid glass substrates using adhesive tape. The strips were placed in quartz EPR tubes which were sealed in a nitrogen glovebox with a bi-component resin (Devcon 5-Minute Epoxy), ensuring that all EPR measurements were performed without oxygen exposure.

All trEPR spectra were recorded on a Bruker Elexsys E580 X-band spectrometer, equipped with a nitrogen gas-flow cryostat for sample temperature control. The sample temperature was maintained with an Oxford Instruments CF935O cryostat and controlled with an Oxford Instruments ITC503. Laser pulses for trEPR were collimated into the cryostat and resonator windows from a multi-mode optical fibre, ThorLabs FT600UMT. Sample excitation at 532 nm with an energy of 2 mJ per pulse and a duration of 7 ns was provided by the residual 2$^{nd}$ harmonic output of a Newport/Spectra Physics Lab 170 Quanta Ray Nd:YAG pulsed laser, operating at 20 Hz. The trEPR signal was recorded through a Bruker SpecJet II transient recorder with timing synchronisation by a Stanford Research Systems DG645 delay generator.

The instrument response time was about 200 ns. The spectra were acquired with 2 mW microwave power and averaging 400 transient signals at each field position.

From the datasets obtained, the transient EPR spectra at different time delays after the laser pulse have been extracted and averaged over a time window of 0.5 μs. Spectral simulations have been performed using the core functions *pepper* and *esfit* of the open-source MATLAB toolbox EasySpin[87]. The parameters included in our best-fit simulations are the ZFS parameters (*D* and *E*), the triplet population sublevels ($p_1$, $p_2$, $p_3$) and the line broadening (assumed as only Lorentzian to not over-parametrize the fitting). For the calculation of spin polarization, the populations of the spin-triplet sublevels at zero field were calculated ($T_x$, $T_y$, $T_z$) in the fitting program and used by EasySpin to simulate the trEPR spectrum at resonant fields. For all the simulations, the *g* tensor was assumed isotropic with $g_{iso}$=2.002. To carry out our least-square fittings, a user-defined simulation function has been developed which allows the fitting of 'non-spin system' parameters, such as the spin populations of the triplet sublevels. All the fits were carried out using a Nelder/Mead downhill simplex optimisation algorithm.

**PLDMR spectroscopy**

PLDMR samples were prepared in the same way as the trEPR samples, whereby the EPR tubes were sealed under inert helium atmosphere to measure at cryogenic (10 K) temperatures. PLDMR measurements were carried out in a modified X-band spectrometer (Bruker E300) equipped with a continuous-flow helium cryostat (Oxford ESR 900) and a microwave cavity (Bruker ER4104OR, 9.43 GHz) with optical access. Optical irradiation was performed with a 473 nm continuous wave laser (Cobolt) from one side opening of the cavity. PL was detected with a silicon photodiode (Hamamatsu Si photodiode S2281) on the opposite opening of the cavity, using a 561 nm longpass filter to reject the excitation wavelength. The PL signal was amplified by a current/voltage amplifier (Femto DHPCA-100) and recorded by lock-in detector (Ametek SR 7230), referenced by TTL-modulating the microwaves with 517 Hz. Microwaves were generated with a microwave signal generator (Anritsu MG3694C), amplified to 3W (Microsemi) and guided into the cavity.

## Data availability

The data that support the plots within this paper is available at the University of Cambridge Repository: [to be completed in proofs].

## Acknowledgements

AJG and RHF acknowledge support from the Simons Foundation (grant no. 601946) and the EPSRC (EP/M01083X/1 and EP/M005143/1). This project has received funding from the ERC under the European Union's Horizon 2020 research and innovation programme (grant agreement no. 670405). AK and T-QN were supported by the Department of the Navy, Office of Naval Research Award No. N00014-21-1-2181. AK acknowledges funding by the Schlumberger foundation. AP, MKR, VD, and AS were supported by the European Union's Horizon 2020 research and innovation programme under Marie Skłodowska Curie grant agreement number 722651 (SEPOMO project). JG, AS, and VD acknowledge support by the Deutsche Forschungsgemeinschaft (DFG, German Research Foundation) within the Research Training School "Molecular biradicals: Structure, properties and reactivity" (GRK2112). trEPR measurements were performed in the Centre for Advanced ESR (CAESR) located in the Department of Chemistry of the University of Oxford, and this work was supported by the EPSRC (EP/L011972/1).

## Author contributions

AJG, AP, JG, and AS conceived the work. AJG performed the TA and PLQE measurements. AP and WKM conducted the trEPR studies. JG carried out the PLDMR measurements. AK fabricated and characterised the OSC devices. VD, T-QN, MKR, RHF, and AS supervised their group members involved in the project. AJG, AP, JG, and AS wrote the manuscript with input from all authors.

## Competing interests

The authors declare no competing interests.

## Additional information

Supplementary information accompanies this paper at [to be completed in proofs].

Correspondence and requests for materials should be addressed to AJG (ajg216@cam.ac.uk) and AS: (sperlich@physik.uni-wuerzburg.de).

## References

1 Q. Liu, Y. Jiang, K. Jin, J. Qin, J. Xu, W. Li, J. Xiong, J. Liu, Z. Xiao, K. Sun, S. Yang, X. Zhang and L. Ding, *Sci. Bull.*, 2020, **65**, 272–275.

2 P. Bi, S. Zhang, Z. Chen, Y. Xu, Y. Cui, T. Zhang, J. Ren, J. Qin, L. Hong, X. Hao and J. Hou, *Joule*, 2021, **5**, 2408–2419.

3 Y. Cai, Y. Li, R. Wang, H. Wu, Z. Chen, J. Zhang, Z. Ma, X. Hao, Y. Zhao, C. Zhang, F. Huang and Y. Sun, *Adv. Mater.*, 2021, **33**, 2101733.

4 Y. Cui, Y. Xu, H. Yao, P. Bi, L. Hong, J. Zhang, Y. Zu, T. Zhang, J. Qin, J. Ren, Z. Chen, C. He, X. Hao, Z. Wei and J. Hou, *Adv. Mater.*, 2021, **33**, 2102420.

5 S. Bao, H. Yang, H. Fan, J. Zhang, Z. Wei, C. Cui and Y. Li, *Adv. Mater.*, 2021, **2105301**, 2105301.

6 S. M. Menke, N. A. Ran, G. C. Bazan and R. H. Friend, *Joule*, 2018, **2**, 25–35.

7 X.-K. Chen, D. Qian, Y. Wang, T. Kirchartz, W. Tress, H. Yao, J. Yuan, M. Hülsbeck, M. Zhang, Y. Zou, Y. Sun, Y. Li, J. Hou, O. Inganäs, V. Coropceanu, J.-L. Bredas and F. Gao, *Nat. Energy*, 2021, **6**, 799–806.

8 S. Liu, J. Yuan, W. Deng, M. Luo, Y. Xie, Q. Liang, Y. Zou, Z. He, H. Wu and Y. Cao, *Nat. Photonics*, 2020, **14**, 300–305.

9 J. Benduhn, K. Tvingstedt, F. Piersimoni, S. Ullbrich, Y. Fan, M. Tropiano, K. A. McGarry, O. Zeika, M. K. Riede, C. J. Douglas, S. Barlow, S. R. Marder, D. Neher, D. Spoltore and K. Vandewal, *Nat. Energy*, 2017, **2**, 17053.

10 W. Shockley and H. J. Queisser, *J. Appl. Phys.*, 1961, **32**, 510–519.

11 R. T. Ross, *J. Chem. Phys.*, 1967, **46**, 4590–4593.

12 U. Rau, *Phys. Rev. B*, 2007, **76**, 085303.

13 B. Geffroy, P. le Roy and C. Prat, *Polym. Int.*, 2006, **55**, 572–582.

14 D. Qian, Z. Zheng, H. Yao, W. Tress, T. R. Hopper, S. Chen, S. Li, J. Liu, S. Chen, J. Zhang, X.-K. Liu, B. Gao, L. Ouyang, Y. Jin, G. Pozina, I. A. Buyanova, W. M. Chen, O. Inganäs, V. Coropceanu, J.-L. Bredas, H. Yan, J. Hou, F. Zhang, A. A. Bakulin and F. Gao, *Nat. Mater.*, 2018, **17**, 703–709.

15 A. Classen, C. L. Chochos, L. Lüer, V. G. Gregoriou, J. Wortmann, A. Osvet, K. Forberich, I. McCulloch, T. Heumüller and C. J. Brabec, *Nat. Energy*, 2020, **5**, 711–719.

16 F. D. Eisner, M. Azzouzi, Z. Fei, X. Hou, T. D. Anthopoulos, T. J. S. Dennis, M. Heeney and J. Nelson, *J. Am. Chem. Soc.*, 2019, **141**, 6362–6374.

17 J. Hou, O. Inganäs, R. H. Friend and F. Gao, *Nat. Mater.*, 2018, **17**, 119–128.

18 R. Englman and J. Jortner, *Mol. Phys.*, 1970, **18**, 145–164.

19 A. Zampetti, A. Minotto and F. Cacialli, *Adv. Funct. Mater.*, 2019, **29**, 1807623.

20 A. J. Gillett, A. Privitera, R. Dilmurat, A. Karki, D. Qian, A. Pershin, G. Londi, W. K. Myers, J. Lee, J. Yuan, S.-J. Ko, M. K. Riede, F. Gao, G. C. Bazan, A. Rao, T.-Q. Nguyen, D. Beljonne and R. H. Friend, *Nature*, 2021, **597**, 666–671.

21 R. Wang, J. Xu, L. Fu, C. Zhang, Q. Li, J. Yao, X. Li, C. Sun, Z.-G. Zhang, X. Wang, Y. Li, J. Ma and M. Xiao, *J. Am. Chem. Soc.*, 2021, **143**, 4359–4366.

22 Z. Chen, X. Chen, Z. Jia, G. Zhou, J. Xu, Y. Wu, X. Xia, X. Li, X. Zhang, C. Deng, Y. Zhang, X. Lu, W. Liu, C. Zhang, Y. (Michael) Yang and H. Zhu, *Joule*, 2021, **5**,

1832–1844.

23 J. M. Marin-Beloqui, D. T. W. Toolan, N. A. Panjwani, S. Limbu, J. Kim and T. M. Clarke, *Adv. Energy Mater.*, 2021, **11**, 2100539.

24 J. Yuan, Y. Zhang, L. Zhou, G. Zhang, H.-L. Yip, T.-K. Lau, X. Lu, C. Zhu, H. Peng, P. A. Johnson, M. Leclerc, Y. Cao, J. Ulanski, Y. Li and Y. Zou, *Joule*, 2019, **3**, 1140–1151.

25 S. M. Menke, A. Sadhanala, M. Nikolka, N. A. Ran, M. K. Ravva, S. Abdel-Azeim, H. L. Stern, M. Wang, H. Sirringhaus, T.-Q. Nguyen, J.-L. Brédas, G. C. Bazan and R. H. Friend, *ACS Nano*, 2016, **10**, 10736–10744.

26 A. Rao, P. C. Y. Chow, S. Gélinas, C. W. Schlenker, C.-Z. Li, H.-L. Yip, A. K.-Y. Jen, D. S. Ginger and R. H. Friend, *Nature*, 2013, **500**, 435–439.

27 I. Ramirez, A. Privitera, S. Karuthedath, A. Jungbluth, J. Benduhn, A. Sperlich, D. Spoltore, K. Vandewal, F. Laquai and M. Riede, *Nat. Commun.*, 2021, **12**, 471.

28 E. M. Speller, A. J. Clarke, N. Aristidou, M. F. Wyatt, L. Francàs, G. Fish, H. Cha, H. K. H. Lee, J. Luke, A. Wadsworth, A. D. Evans, I. McCulloch, J.-S. Kim, S. A. Haque, J. R. Durrant, S. D. Dimitrov, W. C. Tsoi and Z. Li, *ACS Energy Lett.*, 2019, **4**, 846–852.

29 E. M. Speller, J. D. McGettrick, B. Rice, A. M. Telford, H. K. H. Lee, C.-H. Tan, C. S. De Castro, M. L. Davies, T. M. Watson, J. Nelson, J. R. Durrant, Z. Li and W. C. Tsoi, *ACS Appl. Mater. Interfaces*, 2017, **9**, 22739–22747.

30 I. Sudakov, M. Van Landeghem, R. Lenaerts, W. Maes, S. Van Doorslaer and E. Goovaerts, *Adv. Energy Mater.*, 2020, **10**, 2002095.

31 K. Vandewal, S. Mertens, J. Benduhn and Q. Liu, *J. Phys. Chem. Lett.*, 2020, **11**, 129–

135.

32 V. Coropceanu, X.-K. Chen, T. Wang, Z. Zheng and J.-L. Brédas, *Nat. Rev. Mater.*, 2019, **4**, 689–707.

33 A. Karki, A. J. Gillett, R. H. Friend and T.-Q. Nguyen, *Adv. Energy Mater.*, 2021, **11**, 2003441.

34 C. W. Schlenker, K.-S. Chen, H.-L. Yip, C.-Z. Li, L. R. Bradshaw, S. T. Ochsenbein, F. Ding, X. S. Li, D. R. Gamelin, A. K.-Y. Jen and D. S. Ginger, *J. Am. Chem. Soc.*, 2012, **134**, 19661–19668.

35 A. E. Cohen, *J. Phys. Chem. A*, 2009, **113**, 11084–11092.

36 A. J. Gillett, C. Tonnelé, G. Londi, G. Ricci, M. Catherin, D. M. L. Unson, D. Casanova, F. Castet, Y. Olivier, W. M. Chen, E. Zaborova, E. W. Evans, B. H. Drummond, P. J. Conaghan, L.-S. Cui, N. C. Greenham, Y. Puttisong, F. Fages, D. Beljonne and R. H. Friend, *http://arxiv.org/abs/2106.15512*.

37 J. Benduhn, F. Piersimoni, G. Londi, A. Kirch, J. Widmer, C. Koerner, D. Beljonne, D. Neher, D. Spoltore and K. Vandewal, *Adv. Energy Mater.*, 2018, **8**, 1800451.

38 A. Köhler and D. Beljonne, *Adv. Funct. Mater.*, 2004, **14**, 11–18.

39 M.-A. Pan, T.-K. Lau, Y. Tang, Y.-C. Wu, T. Liu, K. Li, M.-C. Chen, X. Lu, W. Ma and C. Zhan, *J. Mater. Chem. A*, 2019, **7**, 20713–20722.

40 M. Zhang, L. Zhu, G. Zhou, T. Hao, C. Qiu, Z. Zhao, Q. Hu, B. W. Larson, H. Zhu, Z. Ma, Z. Tang, W. Feng, Y. Zhang, T. P. Russell and F. Liu, *Nat. Commun.*, 2021, **12**, 309.

41 A. A. Bakulin, S. D. Dimitrov, A. Rao, P. C. Y. Chow, C. B. Nielsen, B. C. Schroeder,

I. McCulloch, H. J. Bakker, J. R. Durrant and R. H. Friend, *J. Phys. Chem. Lett.*, 2013, **4**, 209–215.

42 F. Kraffert, R. Steyrleuthner, S. Albrecht, D. Neher, M. C. Scharber, R. Bittl and J. Behrends, *J. Phys. Chem. C*, 2014, **118**, 28482–28493.

43 M. S. Kotova, G. Londi, J. Junker, S. Dietz, A. Privitera, K. Tvingstedt, D. Beljonne, A. Sperlich and V. Dyakonov, *Mater. Horizons*, 2020, **7**, 1641–1649.

44 J. Lee, S.-J. Ko, M. Seifrid, H. Lee, C. McDowell, B. R. Luginbuhl, A. Karki, K. Cho, T.-Q. Nguyen and G. C. Bazan, *Adv. Energy Mater.*, 2018, **8**, 1801209.

45 H. Yao, Y. Cui, R. Yu, B. Gao, H. Zhang and J. Hou, *Angew. Chemie*, 2017, **129**, 3091–3095.

46 C. Li, J. Zhou, J. Song, J. Xu, H. Zhang, X. Zhang, J. Guo, L. Zhu, D. Wei, G. Han, J. Min, Y. Zhang, Z. Xie, Y. Yi, H. Yan, F. Gao, F. Liu and Y. Sun, *Nat. Energy*, 2021, **6**, 605–613.

47 S.-H. Liao, H.-J. Jhuo, Y.-S. Cheng and S.-A. Chen, *Adv. Mater.*, 2013, **25**, 4766–4771.

48 A. Karki, J. Vollbrecht, A. J. Gillett, P. Selter, J. Lee, Z. Peng, N. Schopp, A. L. Dixon, M. Schrock, V. Nádaždy, F. Schauer, H. Ade, B. F. Chmelka, G. C. Bazan, R. H. Friend and T. Nguyen, *Adv. Energy Mater.*, 2020, **10**, 2001203.

49 S. Karuthedath, J. Gorenflot, Y. Firdaus, N. Chaturvedi, C. S. P. De Castro, G. T. Harrison, J. I. Khan, A. Markina, A. H. Balawi, T. A. Dela Peña, W. Liu, R.-Z. Liang, A. Sharma, S. H. K. Paleti, W. Zhang, Y. Lin, E. Alarousu, D. H. Anjum, P. M. Beaujuge, S. De Wolf, I. McCulloch, T. D. Anthopoulos, D. Baran, D. Andrienko and F. Laquai, *Nat. Mater.*, 2021, **20**, 378–384.

50 L. Franco, A. Toffoletti, M. Ruzzi, L. Montanari, C. Carati, L. Bonoldi and R. Po', *J. Phys. Chem. C*, 2013, **117**, 1554–1560.

51 S. A. J. Thomson, J. Niklas, K. L. Mardis, C. Mallares, I. D. W. Samuel and O. G. Poluektov, *J. Phys. Chem. C*, 2017, **121**, 22707–22719.

52 A. Karki, J. Vollbrecht, A. J. Gillett, S. S. Xiao, Y. Yang, Z. Peng, N. Schopp, A. L. Dixon, S. Yoon, M. Schrock, H. Ade, G. N. M. Reddy, R. H. Friend and T.-Q. Nguyen, *Energy Environ. Sci.*, 2020, **13**, 3679–3692.

53 A. Karki, J. Vollbrecht, A. J. Gillett, P. Selter, J. Lee, Z. Peng, N. Schopp, A. L. Dixon, M. Schrock, V. Nádaždy, F. Schauer, H. Ade, B. F. Chmelka, G. C. Bazan, R. H. Friend and T. Nguyen, *Adv. Energy Mater.*, 2020, **10**, 2001203.

54 A. Rao, P. C. Y. Chow, S. Gélinas, C. W. Schlenker, C.-Z. Li, H.-L. Yip, A. K. Y. Jen, D. S. Ginger and R. H. Friend, *Nature*, 2013, **500**, 435–439.

55 I. A. Howard, R. Mauer, M. Meister and F. Laquai, *J. Am. Chem. Soc.*, 2010, **132**, 14866–14876.

56 S. D. Dimitrov, S. Wheeler, D. Niedzialek, B. C. Schroeder, H. Utzat, J. M. Frost, J. Yao, A. Gillett, P. S. Tuladhar, I. McCulloch, J. Nelson and J. R. Durrant, *Nat. Commun.*, 2015, **6**, 6501.

57 C. Keiderling, S. Dimitrov and J. R. Durrant, *J. Phys. Chem. C*, 2017, **121**, 14470–14475.

58 P. C. Y. Chow, S. Albert-Seifried, S. Gélinas and R. H. Friend, *Adv. Mater.*, 2014, **26**, 4851–4854.

59 S. M. Menke, A. Sadhanala, M. Nikolka, N. A. Ran, M. K. Ravva, S. Abdel-Azeim, H. L. Stern, M. Wang, H. Sirringhaus, T.-Q. Nguyen, J.-L. Brédas, G. C. Bazan and R. H.

Friend, *ACS Nano*, 2016, **10**, 10736–10744.

60 S. Richert, C. E. Tait and C. R. Timmel, *J. Magn. Reson.*, 2017, **280**, 103–116.

61 J. Niklas, S. Beaupré, M. Leclerc, T. Xu, L. Yu, A. Sperlich, V. Dyakonov and O. G. Poluektov, *J. Phys. Chem. B*, 2015, **119**, 7407–7416.

62 T. Biskup, *Front. Chem.*, , DOI:10.3389/fchem.2019.00010.

63 S. Weber, in *eMagRes*, John Wiley & Sons, Ltd, Chichester, UK, 2017, pp. 255–270.

64 S. A. J. Thomson, J. Niklas, K. L. Mardis, C. Mallares, I. D. W. Samuel and O. G. Poluektov, *J. Phys. Chem. C*, 2017, **121**, 22707–22719.

65 C. Hintze, U. E. Steiner and M. Drescher, *ChemPhysChem*, 2017, **18**, 6–16.

66 P. Kohn, Z. Rong, K. H. Scherer, A. Sepe, M. Sommer, P. Müller-Buschbaum, R. H. Friend, U. Steiner and S. Hüttner, *Macromolecules*, 2013, **46**, 4002–4013.

67 N. D. Treat, A. Varotto, C. J. Takacs, N. Batara, M. Al-Hashimi, M. J. Heeney, A. J. Heeger, F. Wudl, C. J. Hawker and M. L. Chabinyc, *J. Am. Chem. Soc.*, 2012, **134**, 15869–15879.

68 E. Buchaca-Domingo, A. J. Ferguson, F. C. Jamieson, T. McCarthy-Ward, S. Shoaee, J. R. Tumbleston, O. G. Reid, L. Yu, M.-B. Madec, M. Pfannmöller, F. Hermerschmidt, R. R. Schröder, S. E. Watkins, N. Kopidakis, G. Portale, A. Amassian, M. Heeney, H. Ade, G. Rumbles, J. R. Durrant and N. Stingelin, *Mater. Horiz.*, 2014, **1**, 270–279.

69 B. A. Collins, Z. Li, J. R. Tumbleston, E. Gann, C. R. Mcneill and H. Ade, *Adv. Energy Mater.*, 2013, **3**, 65–74.

70 N. C. Cates, R. Gysel, Z. Beiley, C. E. Miller, M. F. Toney, M. Heeney, I. McCulloch

and M. D. McGehee, *Nano Lett.*, 2009, **9**, 4153–4157.

71 D. W. Gehrig, I. A. Howard, S. Sweetnam, T. M. Burke, M. D. McGehee and F. Laquai, *Macromol. Rapid Commun.*, 2015, **36**, 1054–1060.

72 J. A. Bartelt, Z. M. Beiley, E. T. Hoke, W. R. Mateker, J. D. Douglas, B. A. Collins, J. R. Tumbleston, K. R. Graham, A. Amassian, H. Ade, J. M. J. Fréchet, M. F. Toney and M. D. McGehee, *Adv. Energy Mater.*, 2013, **3**, 364–374.

73 E. Salvadori, N. Luke, J. Shaikh, A. Leventis, H. Bronstein, C. W. M. Kay and T. M. Clarke, *J. Mater. Chem. A*, 2017, **5**, 24335–24343.

74 A. Karki, J. Vollbrecht, A. L. Dixon, N. Schopp, M. Schrock, G. N. M. Reddy and T.-Q. Nguyen, *Adv. Mater.*, 2019, **31**, 1903868.

75 G. Zhou, H. Ding, L. Zhu, C. Qiu, M. Zhang, T. Hao, W. Feng, Y. Zhang, H. Zhu and F. Liu, *J. Energy Chem.*, 2020, **47**, 180–187.

76 X. Shi, L. Zuo, S. B. Jo, K. Gao, F. Lin, F. Liu and A. K. Y. Jen, *Chem. Mater.*, 2017, **29**, 8369–8376.

77 L. Ye, H. Hu, M. Ghasemi, T. Wang, B. A. Collins, J.-H. Kim, K. Jiang, J. H. Carpenter, H. Li, Z. Li, T. McAfee, J. Zhao, X. Chen, J. L. Y. Lai, T. Ma, J.-L. Bredas, H. Yan and H. Ade, *Nat. Mater.*, 2018, **17**, 253–260.

78 D. Di Nuzzo, A. Aguirre, M. Shahid, V. S. Gevaerts, S. C. J. Meskers and R. A. J. Janssen, *Adv. Mater.*, 2010, **22**, 4321–4324.

79 D. Carbonera, *Photosynth. Res.*, 2009, **102**, 403–414.

80 J. Grüne, V. Dyakonov and A. Sperlich, *Mater. Horizons*, 2021, **8**, 2569–2575.

81 P. A. Lane, X. Wei, Z. V. Vardeny, J. Partee and J. Shinar, *Phys. Rev. B*, 1996, **53**,

R7580–R7583.

82 S. S. Eaton, K. M. More, B. M. Sawant and G. R. Eaton, *J. Am. Chem. Soc.*, 1983, **105**, 6560–6567.

83 O. G. Poluektov, S. Filippone, N. Martín, A. Sperlich, C. Deibel and V. Dyakonov, *J. Phys. Chem. B*, 2010, **114**, 14426–14429.

84 O. G. Poluektov, J. Niklas, K. L. Mardis, S. Beaupré, M. Leclerc, C. Villegas, S. Erten-Ela, J. L. Delgado, N. Martín, A. Sperlich and V. Dyakonov, *Adv. Energy Mater.*, 2014, **4**, 1301517.

85 C. A. Reed and R. D. Bolskar, *Chem. Rev.*, 2000, **100**, 1075–1120.

86 T. Biskup, M. Sommer, S. Rein, D. L. Meyer, M. Kohlstädt, U. Würfel and S. Weber, *Angew. Chemie Int. Ed.*, 2015, **54**, 7707–7710.

87 S. Stoll and A. Schweiger, *J. Magn. Reson.*, 2006, **178**, 42–55.

88 J. Niklas, K. L. Mardis, B. P. Banks, G. M. Grooms, A. Sperlich, V. Dyakonov, S. Beaupré, M. Leclerc, T. Xu, L. Yu and O. G. Poluektov, *Phys. Chem. Chem. Phys.*, 2013, **15**, 9562.

89 B. A. Collins, Z. Li, J. R. Tumbleston, E. Gann, C. R. McNeill and H. Ade, *Adv. Energy Mater.*, 2013, **3**, 65–74.

90 W. Chen, T. Xu, F. He, W. Wang, C. Wang, J. Strzalka, Y. Liu, J. Wen, D. J. Miller, J. Chen, K. Hong, L. Yu and S. B. Darling, *Nano Lett.*, 2011, **11**, 3707–3713.

91 Y. Firdaus, V. M. Le Corre, S. Karuthedath, W. Liu, A. Markina, W. Huang, S. Chattopadhyay, M. M. Nahid, M. I. Nugraha, Y. Lin, A. Seitkhan, A. Basu, W. Zhang, I. McCulloch, H. Ade, J. Labram, F. Laquai, D. Andrienko, L. J. A. Koster and T. D.

Anthopoulos, *Nat. Commun.*, 2020, **11**, 5220.

92 J. Lee, S.-J. Ko, M. Seifrid, H. Lee, B. R. Luginbuhl, A. Karki, M. Ford, K. Rosenthal, K. Cho, T.-Q. Nguyen and G. C. Bazan, *Adv. Energy Mater.*, 2018, **8**, 1801212.

93 J. C. de Mello, H. F. Wittmann and R. H. Friend, *Adv. Mater.*, 1997, **9**, 230–232.

# Supplementary information for

## Combining optical and magnetic resonance spectroscopies to probe charge recombination via triplet excitons in organic solar cells

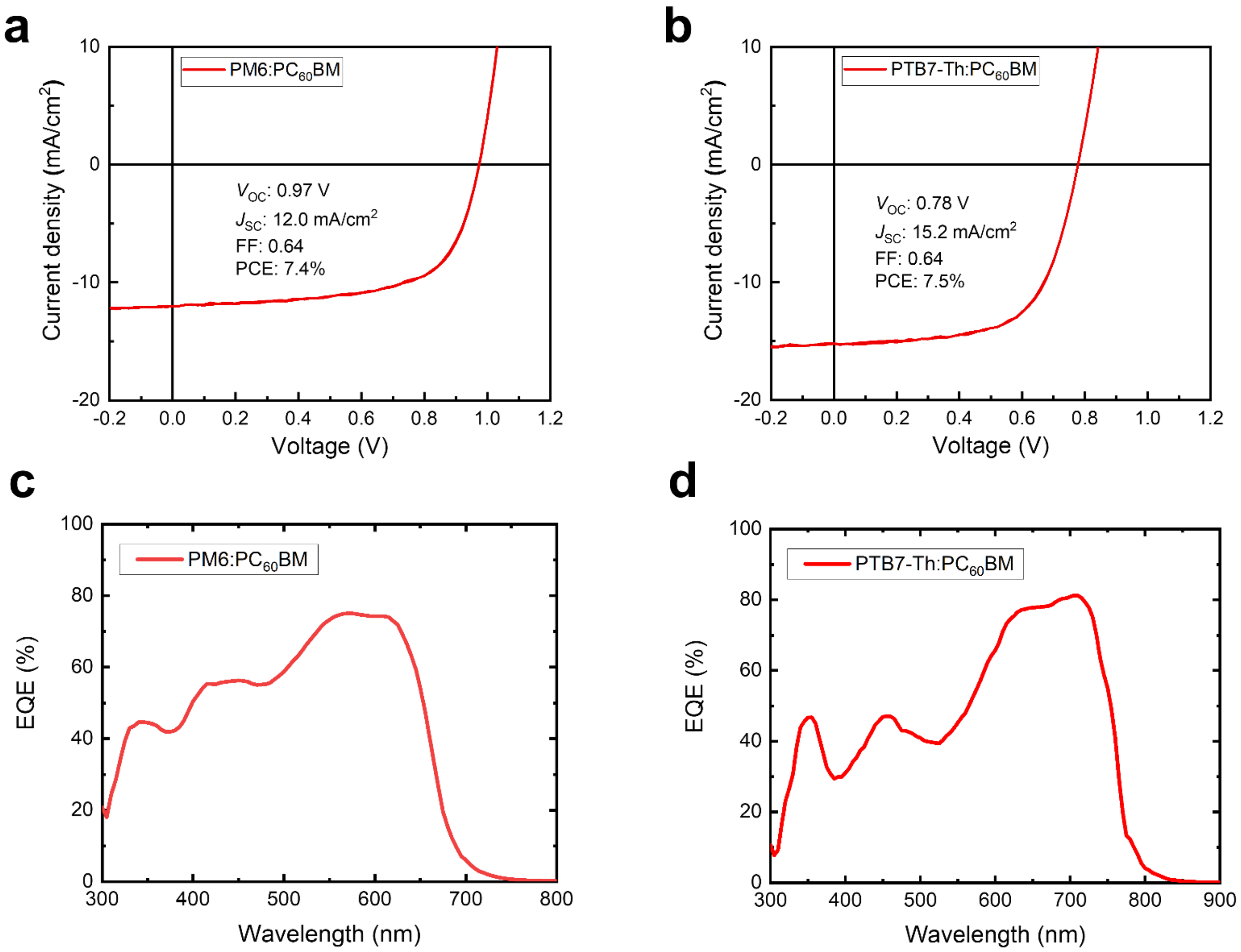

**Figure S1: (a)** The current density-voltage curve of the conventional architecture PM6:$PC_{60}BM$ solar cell studied in this work. **(b)** The current density-voltage curve of the inverted architecture PTB7-Th:$PC_{60}BM$ solar cell studied in this work. **(c)** The external quantum efficiency spectrum of the PM6:$PC_{60}BM$ solar cell studied in this work. **(d)** The external quantum efficiency spectrum of the PTB7-Th:$PC_{60}BM$ solar cell studied in this work.

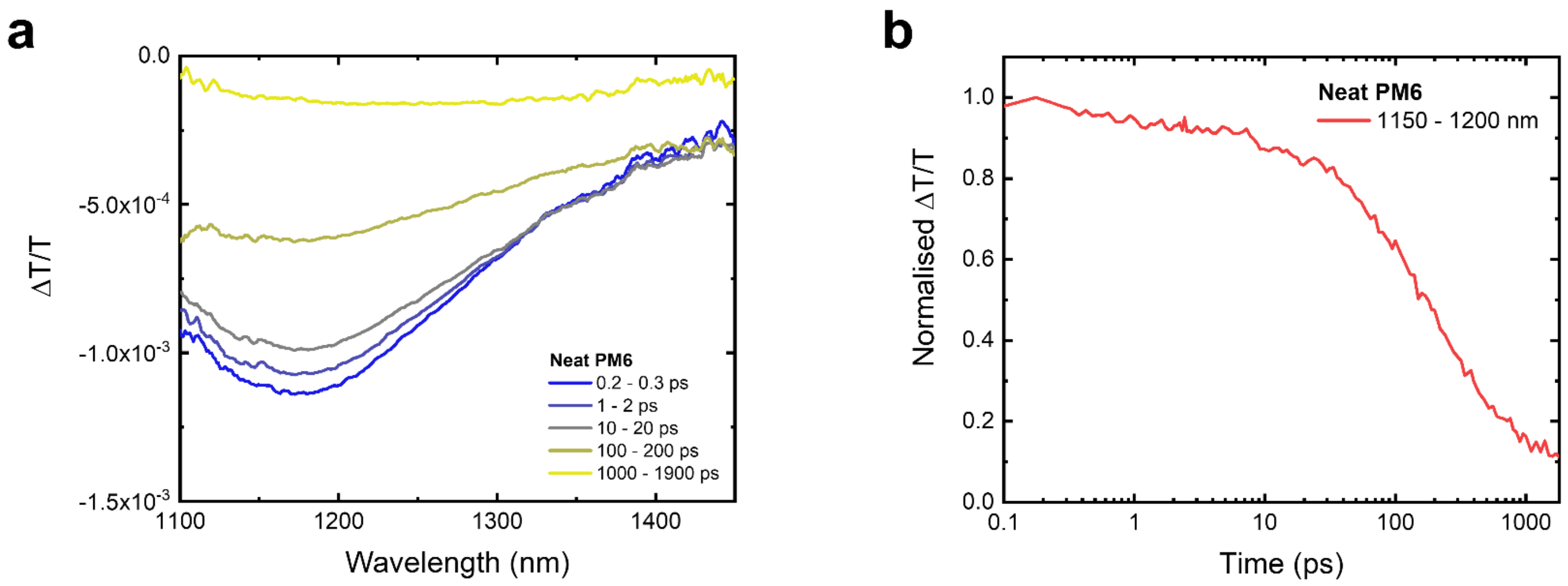

**Figure S2: (a)** The NIR region TA spectra of a neat PM6 film, excited at 600 nm with a fluence of 1.3 μJ $cm^{-2}$. The PM6 $S_1$ PIA can be seen at 1175 nm. **(b)** The corresponding TA kinetics of the neat PM6 film, taken around the maximum of the PM6 $S_1$ PIA between 1150-1200 nm.

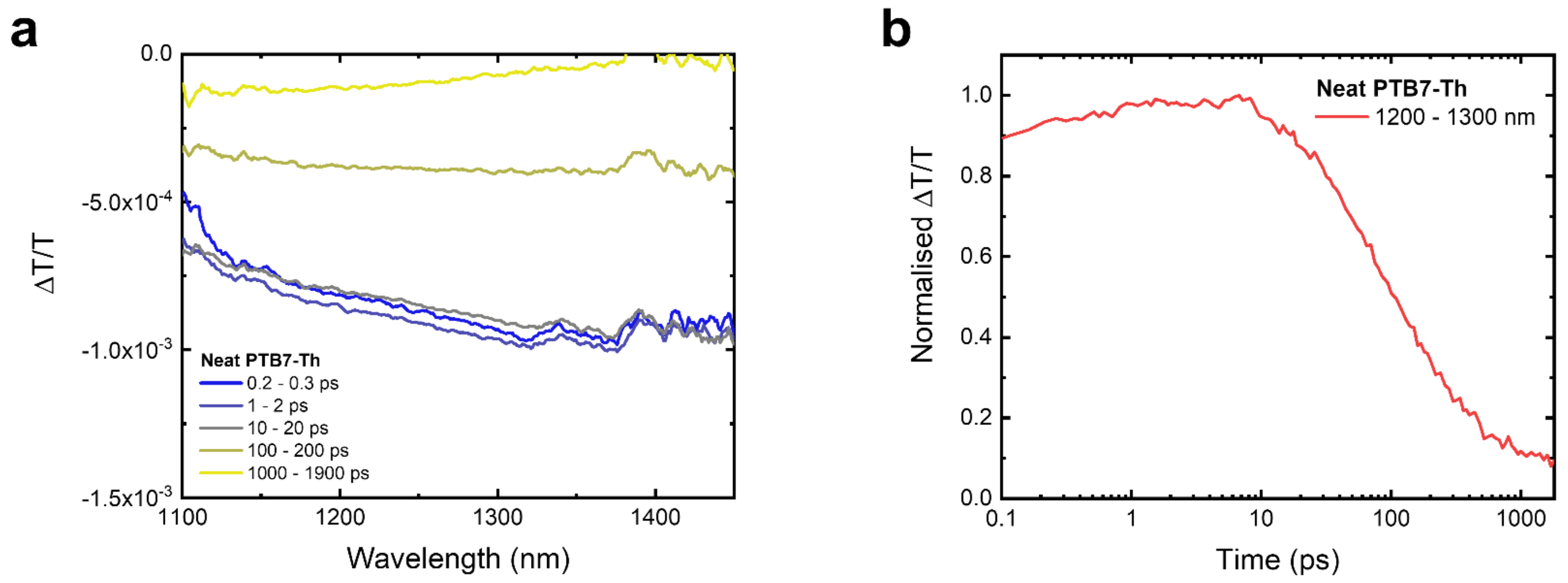

**Figure S3: (a)** The NIR region TA spectra of a neat PTB7-Th film, excited at 700 nm with a fluence of 1.9 μJ $cm^{-2}$. The PTB7-Th $S_1$ PIA is a broad band spanning the NIR probe region between 1100-1450 nm. **(b)** The corresponding TA kinetics of the neat PTB7-Th film, taken from the $S_1$ PIA between 1200-1300 nm.

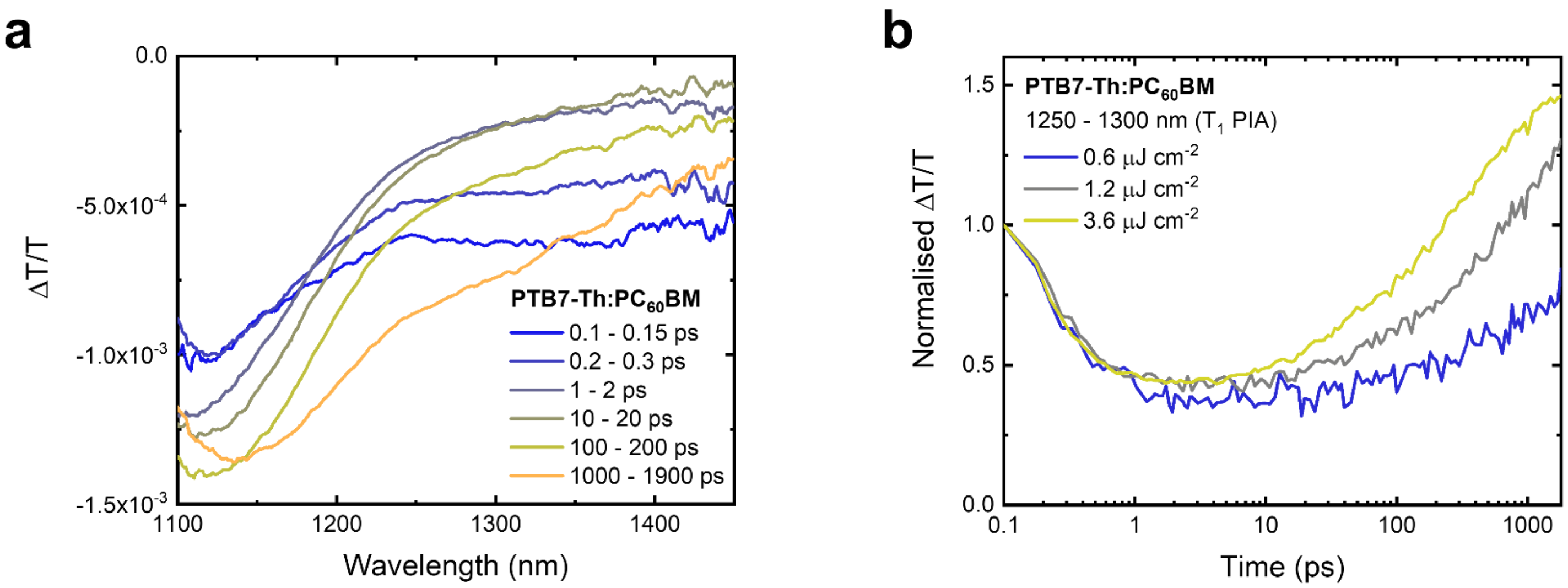

**Figure S4: (a)** The TA spectra of a PTB7-Th:$PC_{60}BM$ blend film, excited at 700 nm with a fluence of 3.6 μJ $cm^{-2}$. The PTB7-Th $S_1$ PIA band around 1400 nm decays within the first picosecond due to electron transfer to $PC_{60}BM$, leaving behind the PTB7-Th hole polaron PIA at 1125 nm. Over hundreds of picoseconds, a new PIA band on the low energy edge of the hole polaron around 1300 nm begins to grow in, indicating recombination into PTB7-Th $T_1$ states. **(b)** The TA kinetics a PTB7-Th:$PC_{60}BM$ blend film, excited at 700 nm with varying fluence. The fluence dependence of the $T_1$ PIA region growth shows that $T_1$ formation occurs following the bimolecular recombination of free charge carriers.

| Film | [D E] /MHz | Triplet? | ISC triplet | | | Geminate BCT triplet | | | NRMSD | Charges? |
|---|---|---|---|---|---|---|---|---|---|---|
| | | | $[p_1\ p_2\ p_3]$ | LW /mT | Weight | $[p_{+1}\ p_0\ p_{-1}]$ | LW /mT | Weight | | |
| $PC_{60}BM$ (1 µs) | [-237 39] | A | [0.34 0.40 0.26] | 1.59 | 0.66 | [0 1 0] | 2.08 | 0.34 | 0.02 | Weak FC |
| $PC_{60}BM$ (5 µs) | [-256 35] | A | [0.40 0.38 0.22] | 2.04 | 0.78 | [0 1 0] | 3.11 | 0.22 | 0.01 | Weak FC |
| PM6 (1 µs) | [1410 125] | D | [0.28 0.36 0.36] | 8 | | | | | 0.08 | FC |
| PM6 (5 µs) | [135 0] | D | [0.34 0.43 0.23] | 5 | | | | | 0.10 | FC |
| PTB7-Th (1 µs) | [1100 160] | D | [0.12 0.43 0.45] | 11 | | | | | 0.03 | FC |
| PTB7-Th (5 µs) | [1050 160] | D | [0.47 0.28 0.25] | 8 | | | | | 0.04 | FC |
| PM6:$PC_{60}BM$ (1 µs) | [1300 140] | D | [0.14 0.52 0.34] | 12 | | | | | 0.16 | CT |
| PM6:$PC_{60}BM$ (5 µs) | [1220 40] | D | [0.39 0.2 0.41] | 13 | 0.57 | [0 1 0] | 5 | 0.43 | 0.04 | FC |
| PTB7-Th:$PC_{60}BM$ (1 µs) | [1050 200] | D | [0.37 0.33 0.30] | 9 | 0.63 | [0 1 0] | 13 | 0.37 | 0.03 | CT |
| PTB7-Th:$PC_{60}BM$ (5 µs) | [1143 164] | D | [0.35 0.33 0.32] | 7 | 0.78 | [0 1 0] | 11 | 0.22 | 0.04 | CT |

**Table S1: Summary of the best-fit spectral simulations of the trEPR measurements reported in Figure 2.** For each sample, the ZFS parameters of the triplet states, given in units of MHz, are reported. From the ZFS parameters, we assigned the triplet either to the donor (D) or the acceptor (A). Two different populating mechanisms have been considered to fit the trEPR spectra: SOC-promoted ISC and geminate BCT. For ISC, the populations are ordered from low-to-high energy zero-field states, $T_z$, $T_x$, and $T_y$, respectively, for D > 0 & E > 0. For BCT, the populations of the high-field levels ($T_+$, $T_0$ and $T_-$) are reported. Only Lorentzian broadening was considered to avoid over-parametrizing the fitting; the linewidth is reported in units of mT. The normalised root-mean-square-deviation (NRMSD) is also reported. Finally, the presence of charges (either a CT state or free charges, FC) is summarized.

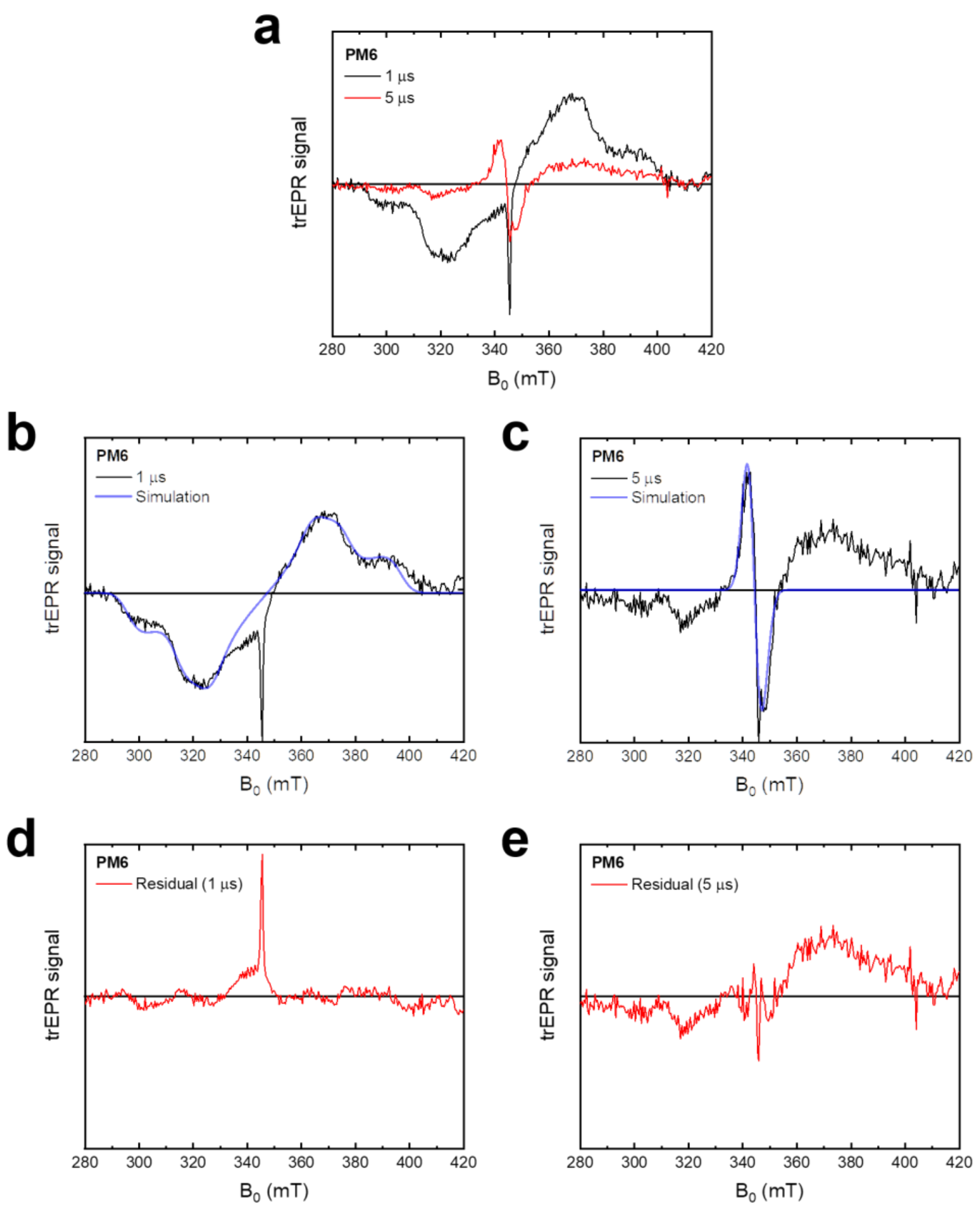

**Figure S5: (a)** The trEPR spectra of a neat PM6 film, taken at representative time points of 1 and 5 μs after excitation at 532 nm. Absorption (*a*) is up, emission (*e*) is down. **(b)** The trEPR spectra at 1 μs is shown, with the simulation overlaid. **(c)** The trEPR spectra at 5 μs is shown, with the simulation of the central, narrower triplet feature between 330 – 360 mT overlaid. **(d)** The residual from the best fit simulation in Figure S5b, shown on the same y-axis scale. The residual (excluding the polaron region, which was not included in the simulation) indicates that the simulation describes the experimental spectrum well. **(e)** The residual from the best fit simulation in Figure S5c, shown on the same y-axis scale. The residual in the 330 – 360 mT region of the narrower triplet feature (excluding the polaron region, which was not included in the simulation) indicates that the simulation describes the experimental spectrum well. The remaining residual is due to the presence of a more localised triplet exciton with a larger *D* parameter, likely the triplet observed at 1 μs in Figure S5a.

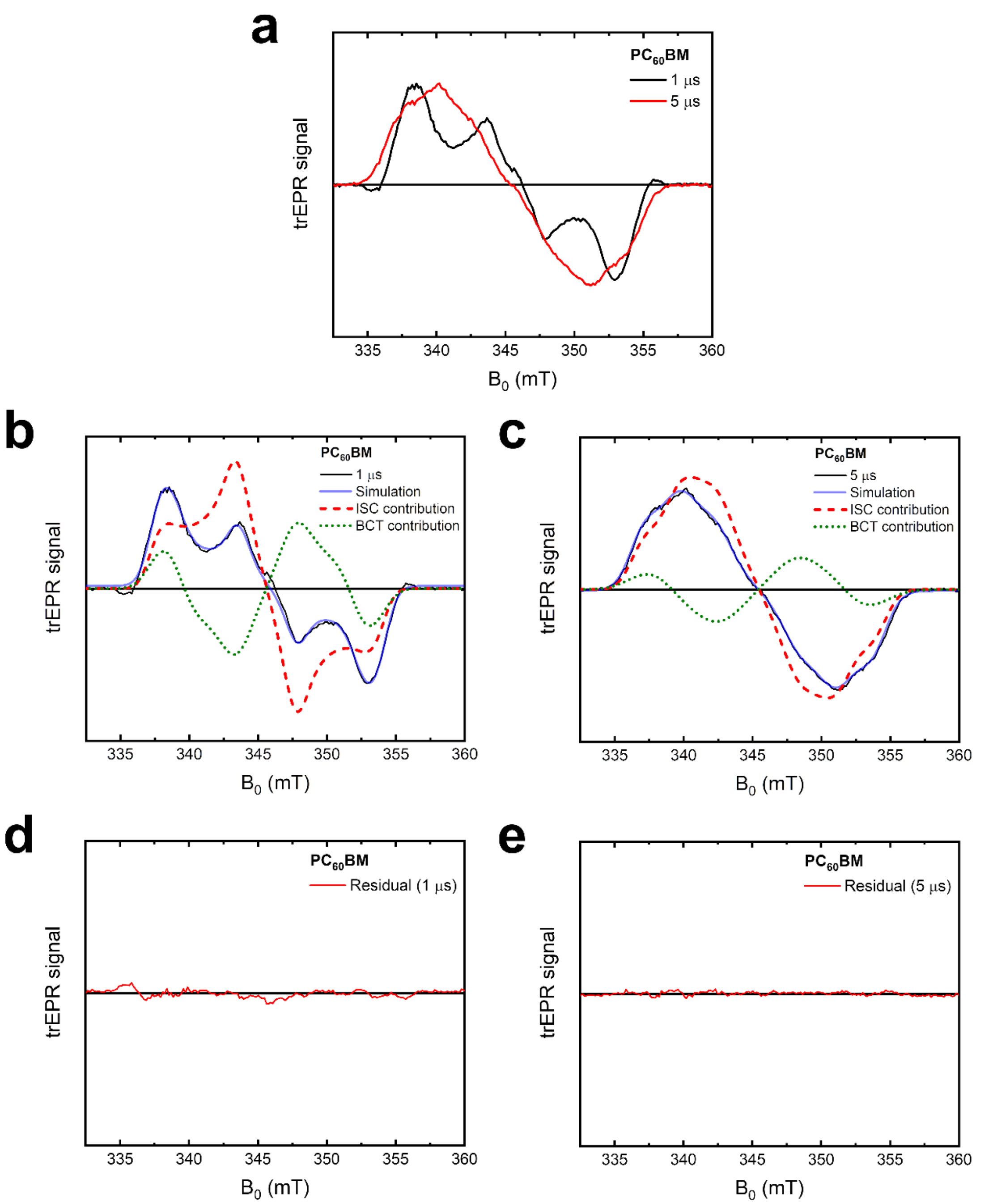

**Figure S6: (a)** The trEPR spectra of a neat $PC_{60}BM$ film, taken at representative time points of 1 and 5 µs after excitation at 532 nm. Absorption (*a*) is up, emission (*e*) is down. **(b)** The trEPR spectra at 1 µs is shown, with the simulation overlaid. **(c)** The trEPR spectra at 5 µs is shown. **(d)** The residual from the best fit simulation in Figure S6b, shown on the same y-axis scale. The residual indicates that the simulation describes the experimental spectrum well. **(e)** The residual from the best fit simulation in Figure S6c, shown on the same y-axis scale. The indicates that the simulation describes the experimental spectrum well.

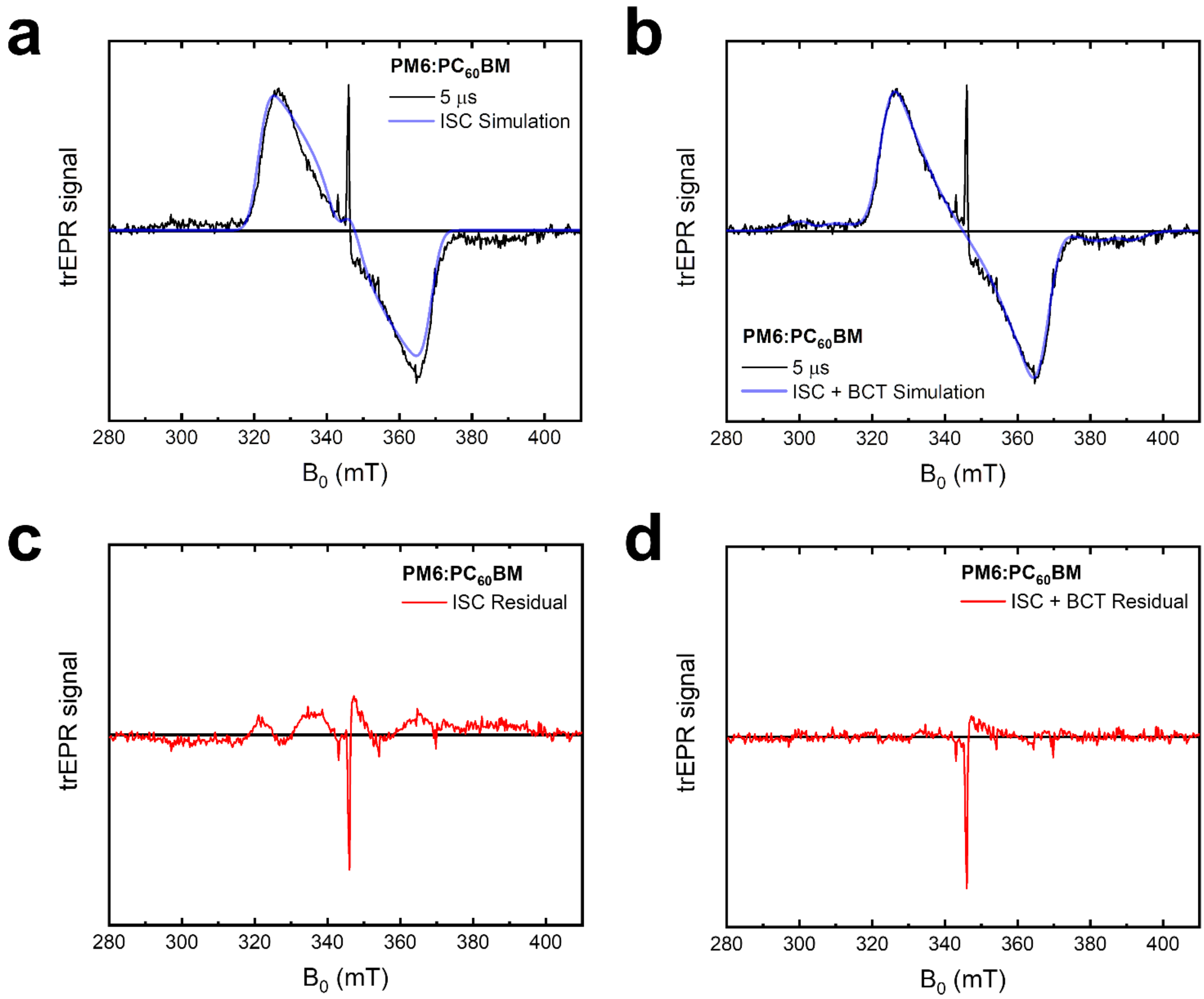

**Figure S7: (a)** The trEPR spectra of a PM6:$PC_{60}$BM film, taken at 5 μs after excitation at 532 nm. Absorption (*a*) is up, emission (*e*) is down. The best fit spectral simulation using a single ISC $T_1$ species is shown. **(b)** The trEPR spectra of a PM6:$PC_{60}$BM film, taken at 5 μs after excitation at 532 nm. Absorption (*a*) is up, emission (*e*) is down. The best fit spectral simulation using both an ISC and BCT $T_1$ species is shown. **(c)** The residual from the best fit ISC simulation in Figure S7a, shown on the same y-axis scale. The single species is not sufficient to fully describe the measured spectrum, as evident from the highly structured residual between 320-370 mT and the spectral 'wings' between 290-320 and 380-410 mT, which cannot be fitted in the ISC-only simulation. **(d)** The residual from the best fit ISC and BCT combined simulation in Figure S7b, shown on the same y-axis scale. By using two species, an excellent agreement is found with the experimental spectrum, indicating the presence of ISC and BCT $T_1$ states in this sample.

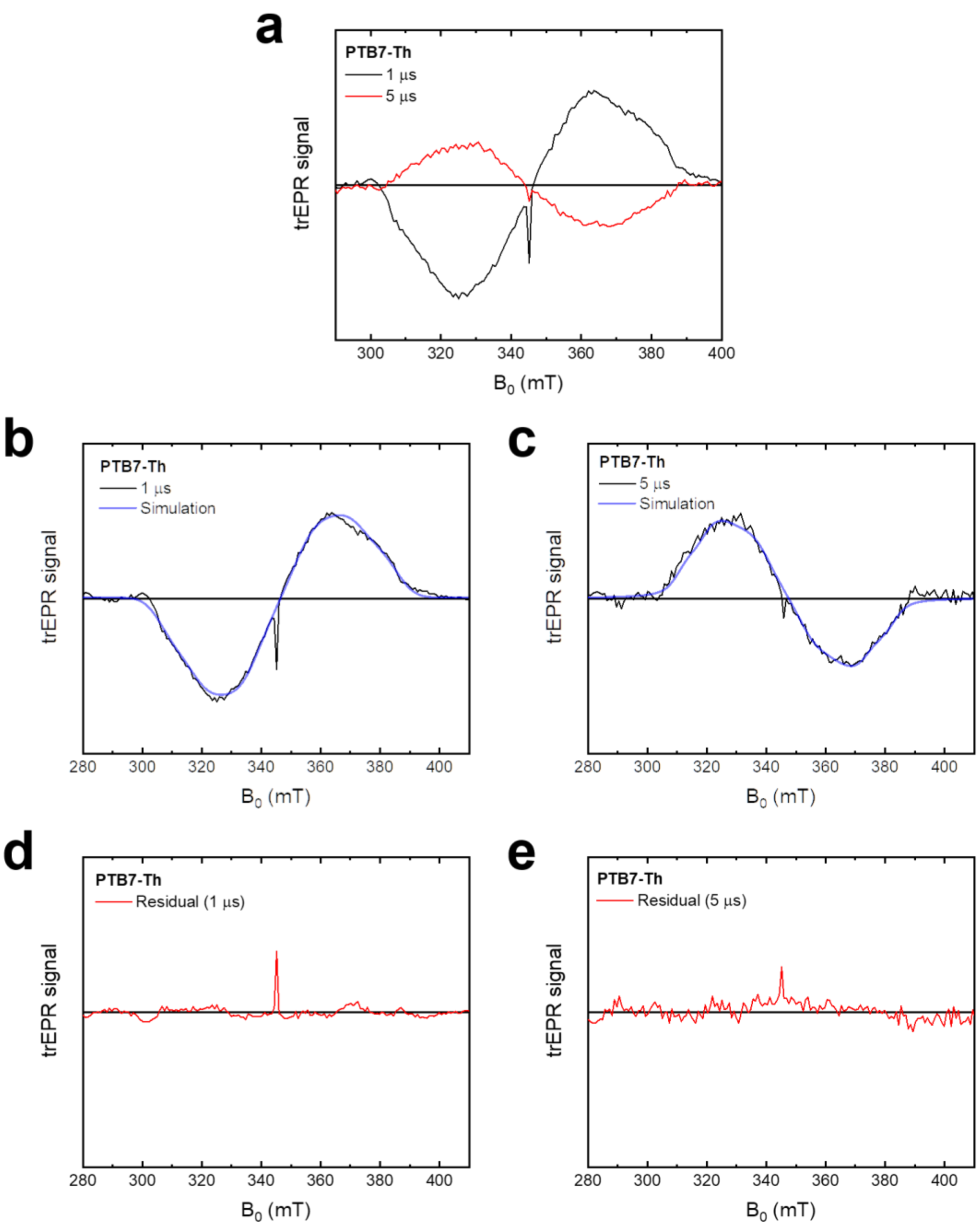

**Figure S8: (a)** The trEPR spectra of a neat PTB7-Th film, taken at representative time points of 1 and 5 μs after excitation at 532 nm. Absorption (*a*) is up, emission (*e*) is down. **(b)** The trEPR spectra at 1 μs is shown, with the simulation overlaid. **(c)** The trEPR spectra at 5 μs is shown, with the simulation overlaid. **(d)** The residual from the best fit simulation in Figure S8b, shown on the same y-axis scale. The residual (excluding the polaron region, which was not included in the simulation) indicates that the simulation describes the experimental spectrum well. **(e)** The residual from the best fit simulation in Figure S8c, shown on the same y-axis scale. The residual (excluding the polaron region, which was not included in the simulation) indicates that the simulation describes the experimental spectrum well.

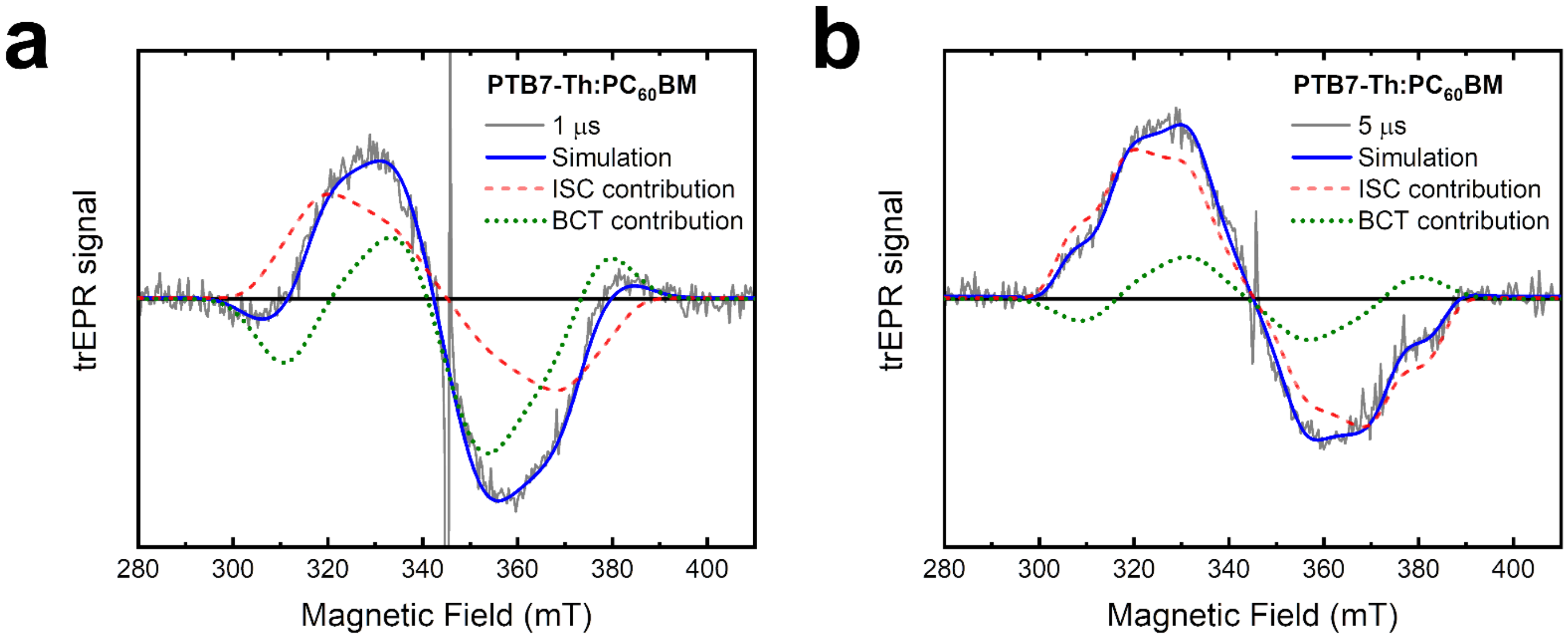

**Figure S9: (a)** The trEPR spectrum of a PTB7-Th:$PC_{60}BM$ blend film at 1 μs, excited at 532 nm. Absorption (*a*) is up, emission (*e*) is down. The central *ea* species is assigned to a CT state. The PTB7-Th $T_1$ species requires the use of two $T_1$ contributions to successfully describe the observed spectrum. The first is an *aaaeee* component, indicating $T_1$ states formed *via* direct ISC from undissociated $S_1$ states. The second is an *eaaeea* component, which is characteristic of $T_1$ states formed *via* the geminate BCT mechanism. **(b)** The trEPR spectrum of a PTB7-Th:$PC_{60}BM$ blend film at 5 μs, excited at 532 nm. The *ea* CT state is still present, indicating slower generation of free charges in this blend. The PTB7-Th $T_1$ species requires the use of two $T_1$ contributions to successfully describe the observed spectrum. The first is an *aaaeee* component, indicating $T_1$ states formed *via* direct ISC from undissociated $S_1$ states. The second is an *eaaeea* component, which is characteristic of $T_1$ states formed *via* the geminate BCT mechanism.

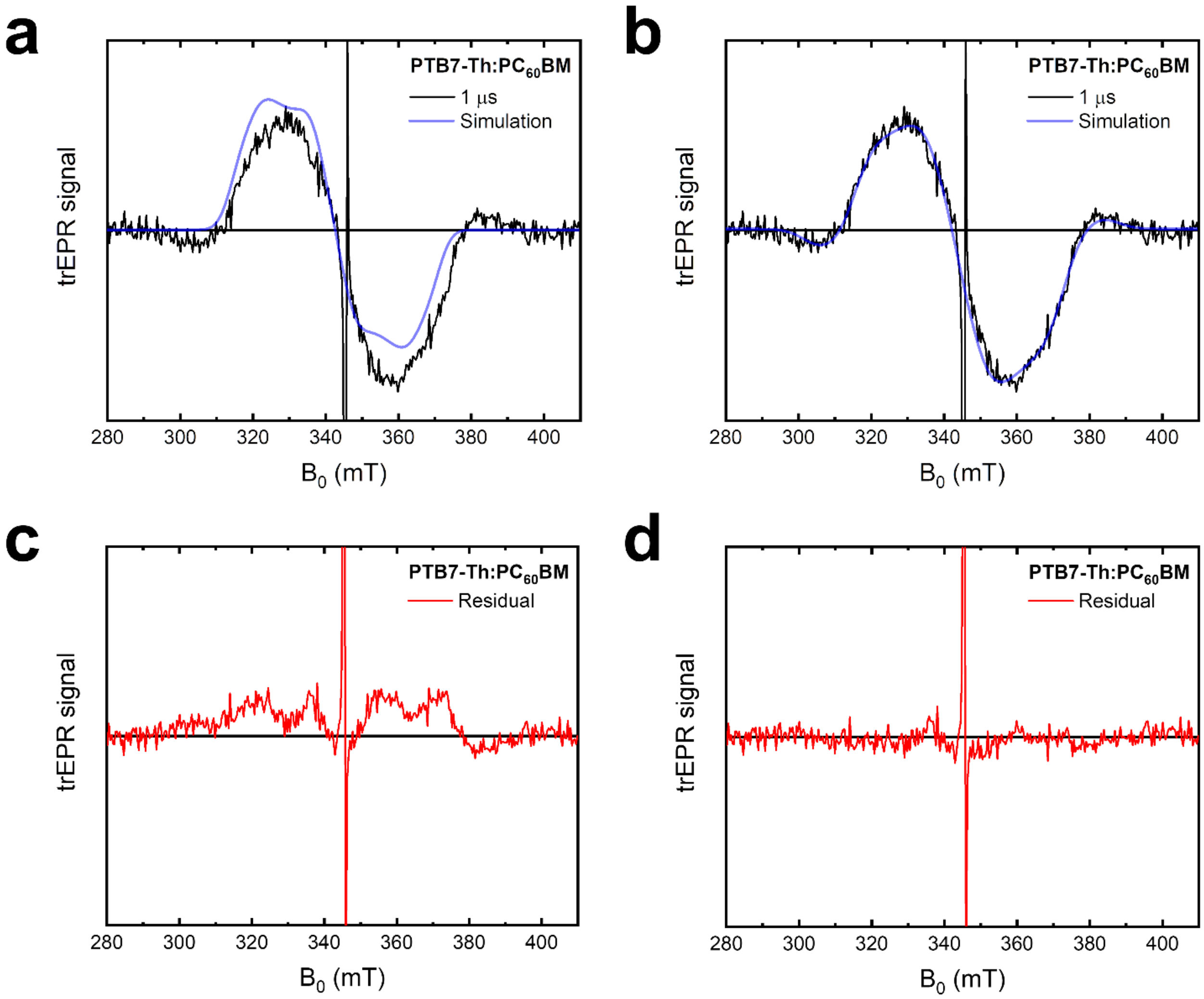

**Figure S10: (a)** The trEPR spectra of a PTB7-Th:$PC_{60}BM$ film, taken at 1 μs after excitation at 532 nm. Absorption (*a*) is up, emission (*e*) is down. The best fit spectral simulation using a single ISC $T_1$ species is shown. **(b)** The trEPR spectra of a PTB7-Th:$PC_{60}BM$ film, taken at 1 μs after excitation at 532 nm. Absorption (*a*) is up, emission (*e*) is down. The best fit spectral simulation using both an ISC and BCT $T_1$ species is shown. **(c)** The residual from the best fit ISC simulation in Figure S10a, shown on the same y-axis scale. The single species is clearly not sufficient to fully describe the measured spectrum, as evident from the highly structured residual. **(d)** The residual from the best fit ISC and BCT combined simulation in Figure S10b, shown on the same y-axis scale. By using two species, an excellent agreement is found with the experimental spectrum, indicating the presence of ISC and BCT $T_1$ states in this sample.

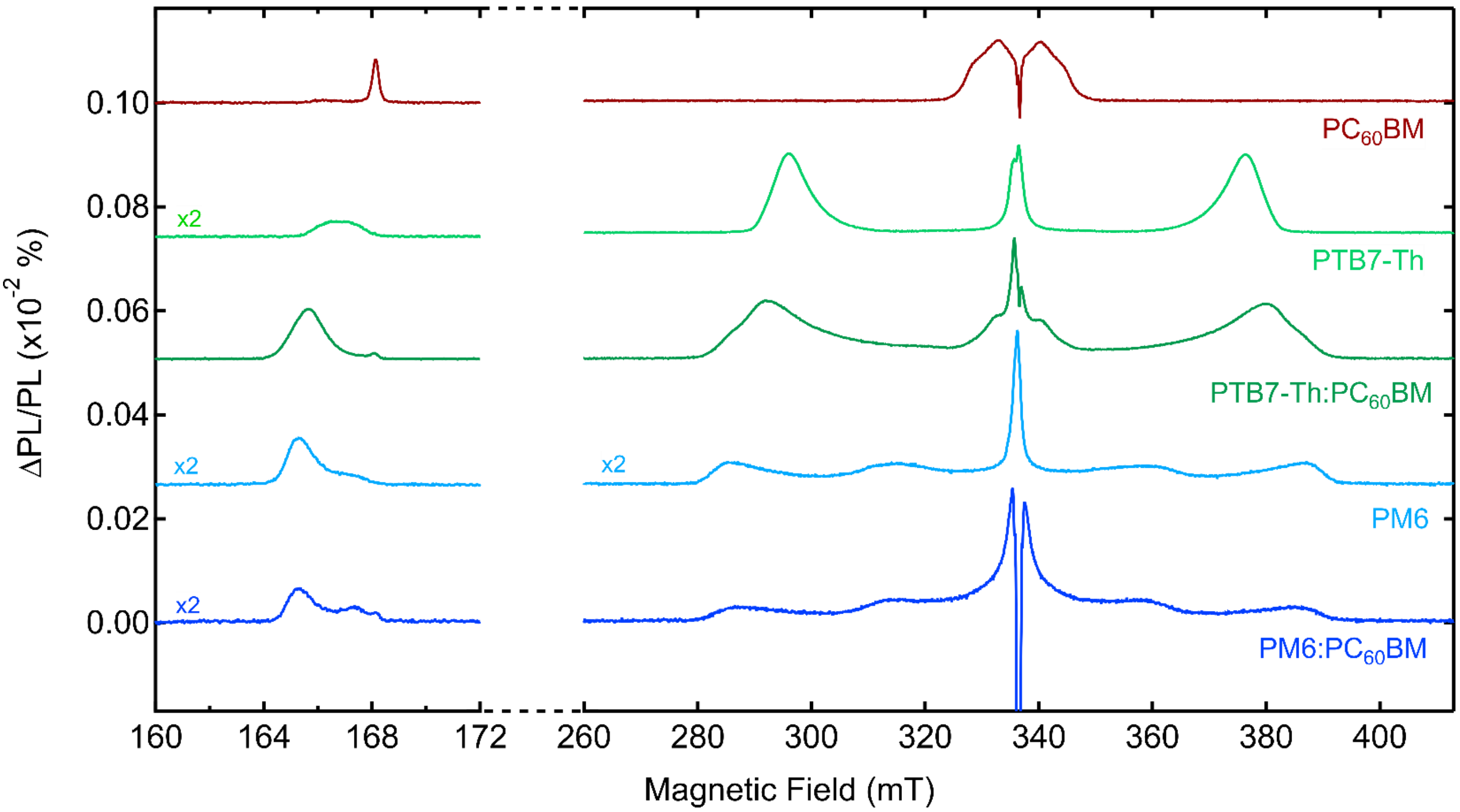

**Figure S11: (Red)** The half-field (160-172 mT) and full-field (260-410 mT) PLDMR spectrum of a neat $PC_{60}BM$ film. The HF $T_1$ signal of $PC_{60}BM$ is visible at 168.1 mT, whilst the FF $T_1$ and polaron features are present between 320-350 mT. **(Light green)** The HF and FF PLDMR spectrum of a neat PTB7-Th film. The HF $T_1$ signal of PTB7-Th is visible at 166.8 mT, whilst the FF $T_1$ manifests as the two spectral 'wings' at 296 and 376 mT. **(Dark green)** The HF and FF PLDMR spectrum of the PTB7-Th:$PC_{60}BM$ blend film. The HF $T_1$ signal of PTB7-Th is visible at 165.6 mT, slightly shifted from the neat film due to the changes in the polymer chain ordering upon blending with $PC_{60}BM$. The HF signal of the $PC_{60}BM$ $T_1$ is weakly visible at 168.1 mT. The FF PTB7-Th $T_1$ is visible between 280-390 mT. **(Light blue)** The HF and FF PLDMR spectrum of a neat PM6 film. The HF $T_1$ signal of PM6 is visible at 165.3 mT, whilst the FF $T_1$ is the broad feature spanning 280-390 mT. **(Dark blue)** The HF and FF PLDMR spectrum of the PM6:$PC_{60}BM$ blend film. The HF $T_1$ signals of PM6 and $PC_{60}BM$ are visible at 165.3 mT and 168.1 mT, respectively. The FF PM6 $T_1$ is visible between 280-390 mT. All PLDMR spectra were acquired at 10 K.

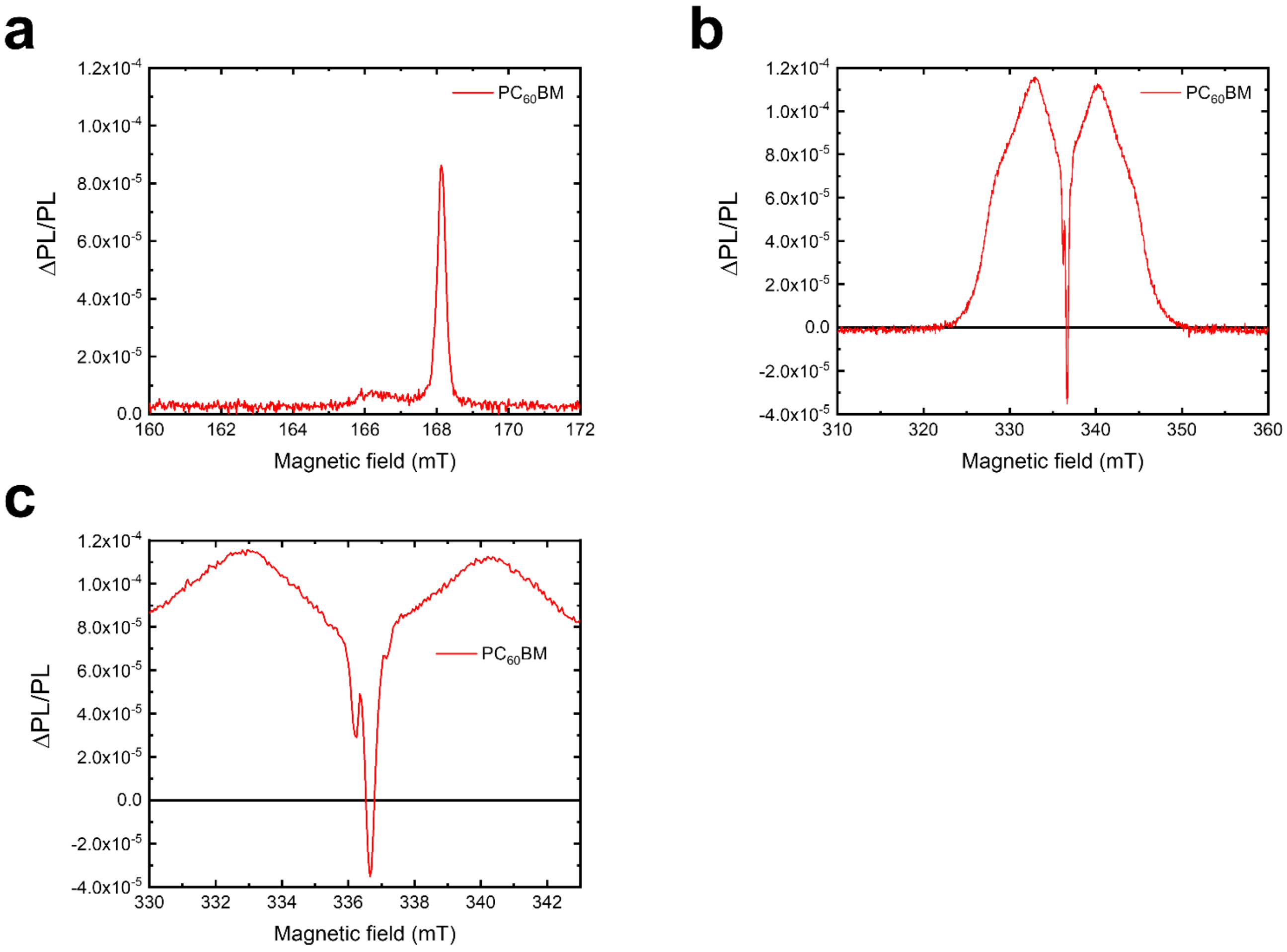

**Figure S12: (a)** The HF PLDMR spectrum of a neat $PC_{60}BM$ film. The HF $PC_{60}BM$ $T_1$ feature appears at 168.1 mT. **(b)** The zoomed in FF PLDMR spectrum of a neat $PC_{60}BM$ film. The FF $PC_{60}BM$ $T_1$ feature spans 320-350 mT. **(c)** The central region of the neat $PC_{60}BM$ film. Two individual polaron features can be seen at 336.25 mT (g = 2.0012) and 336.65 mT (g = 2.0040), corresponding to the $PC_{60}BM$ negative and positive polarons, respectively.

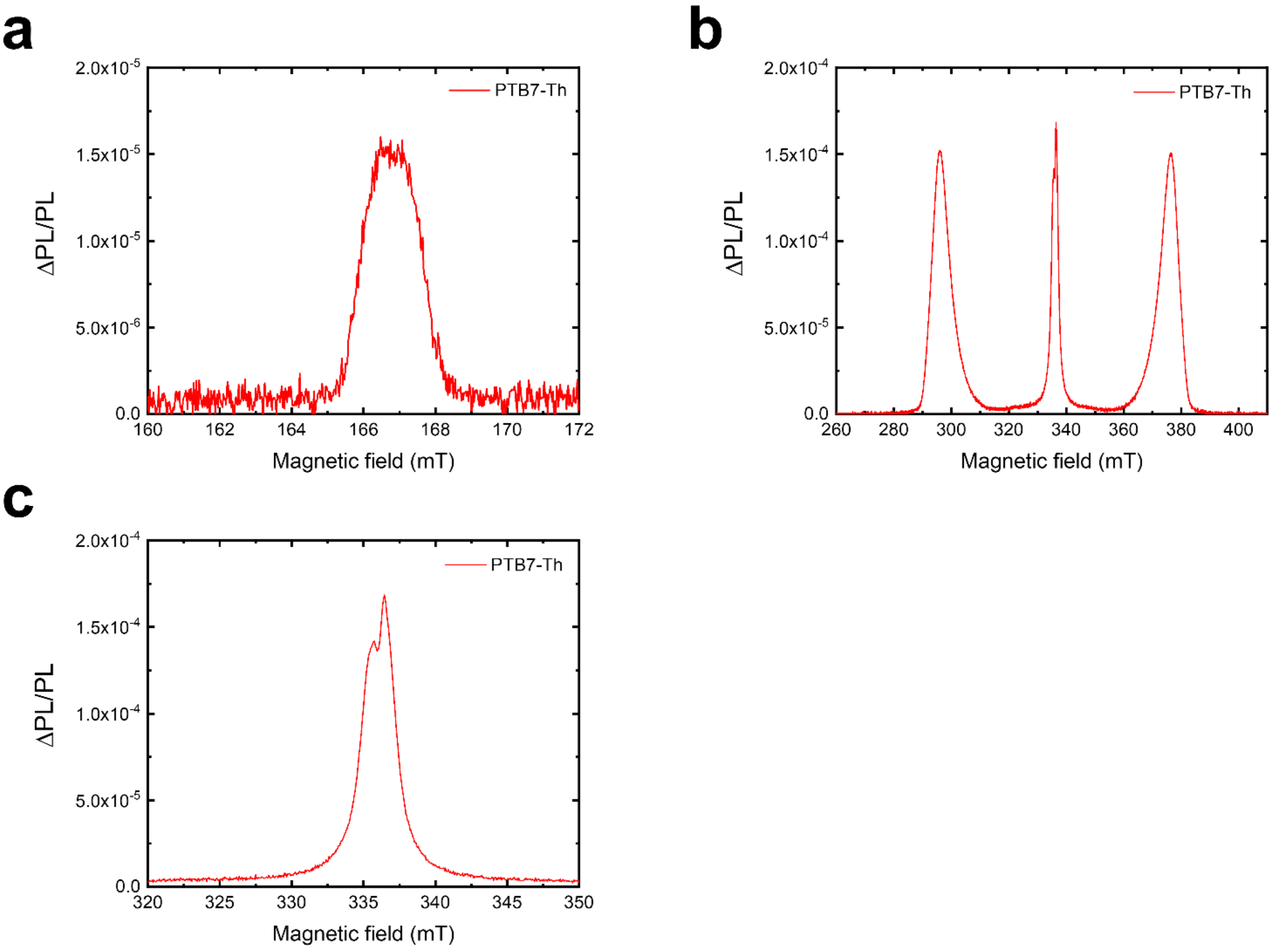

**Figure S13: (a)** The HF PLDMR spectrum of a neat PTB7-Th film. The HF PTB7-Th $T_1$ feature appears at 166.8 mT. **(b)** The zoomed in FF PLDMR spectrum of a neat PTB7-Th film. The FF PTB7-Th $T_1$ feature spans 285-385 mT. **(c)** The central region of the neat PTB7-Th film.

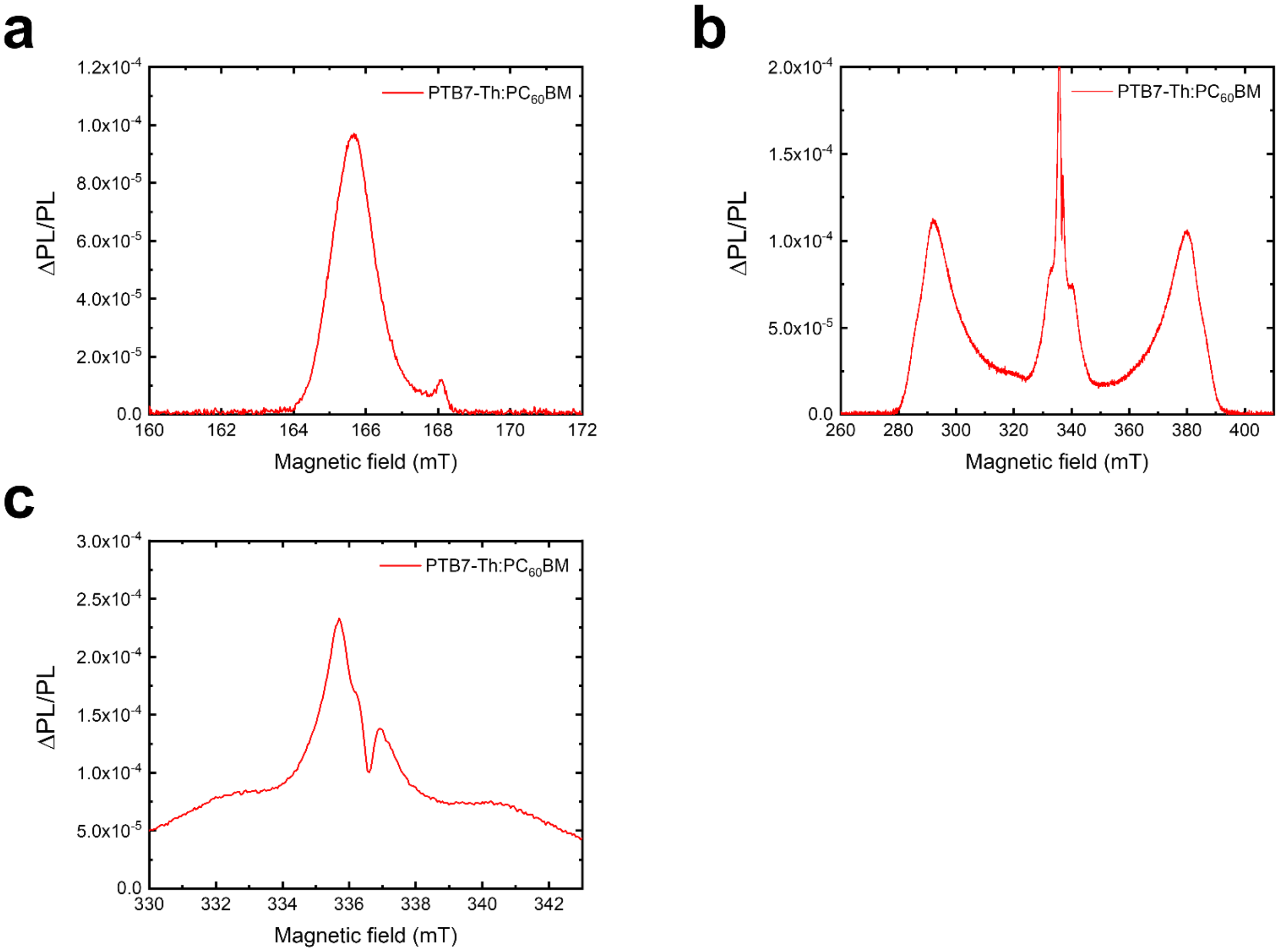

**Figure S14: (a)** The HF PLDMR spectrum of the PTB7-Th:$PC_{60}$BM blend film. The HF PTB7-Th $T_1$ feature appears at 165.6 mT, slightly shifted from the neat PTB7-Th due to a change in the ordering of the polymer chains. The $PC_{60}$BM HF $T_1$ signal is weakly visible at 168.1 mT. **(b)** The zoomed in FF PLDMR spectrum of the PTB7-Th:$PC_{60}$BM blend film. The FF PTB7-Th $T_1$ feature spans 280-390 mT. **(c)** The central region of the PTB7-Th:$PC_{60}$BM blend film. Two polarons with $g = 2.0012$ and $g = 2.0037$ are visible; the former corresponds to the negative $PC_{60}$BM polaron, whilst the latter is the PTB7-Th positive polaron.

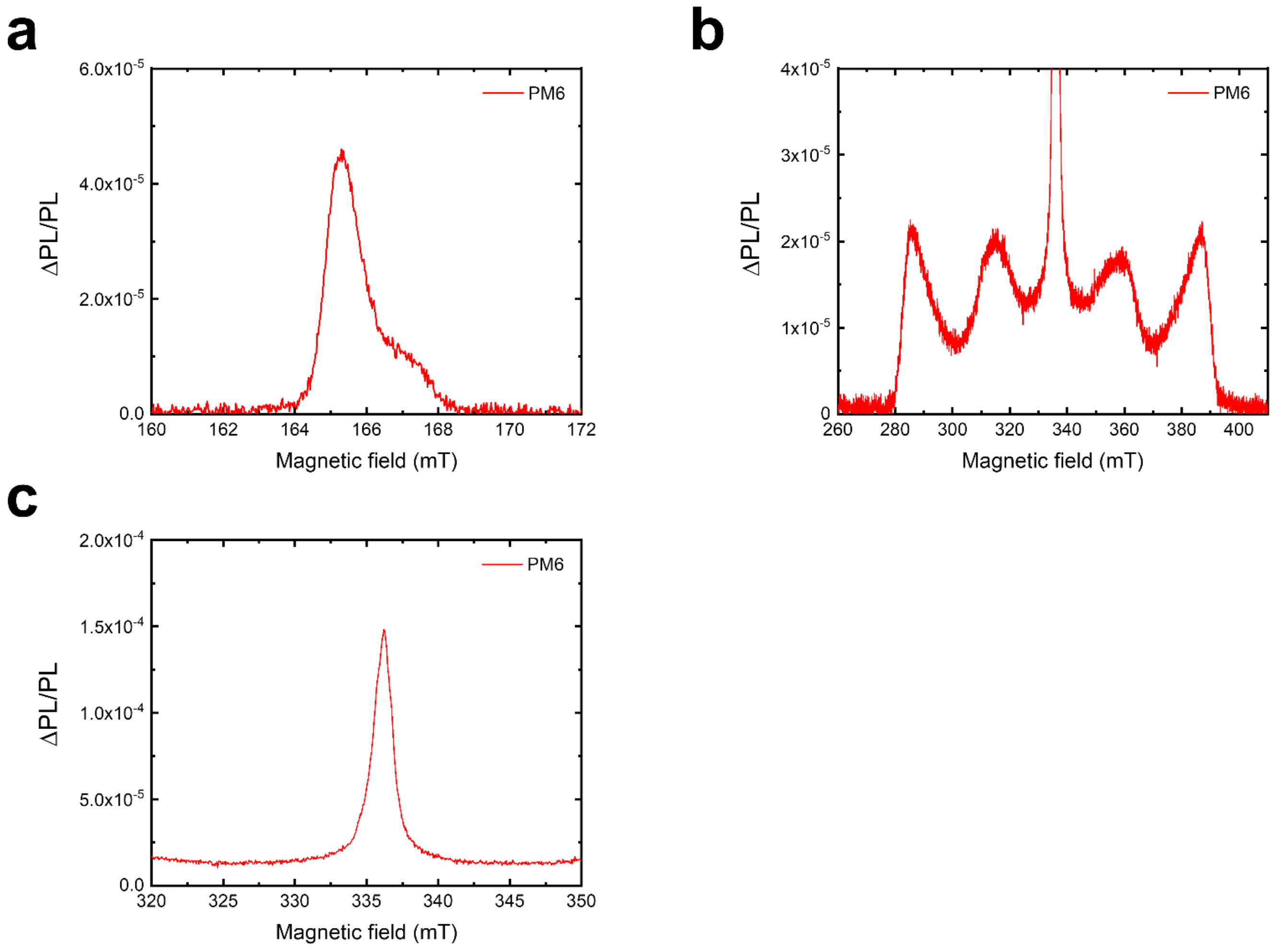

**Figure S15: (a)** The HF PLDMR spectrum of a neat PM6 film. The HF PM6 $T_1$ feature appears at 165.3 mT, with a small shoulder at ~167 mT. **(b)** The zoomed in FF PLDMR spectrum of a neat PM6 film. The FF PM6 $T_1$ feature spans 280-390 mT. **(c)** The central region of the neat PM6 film.

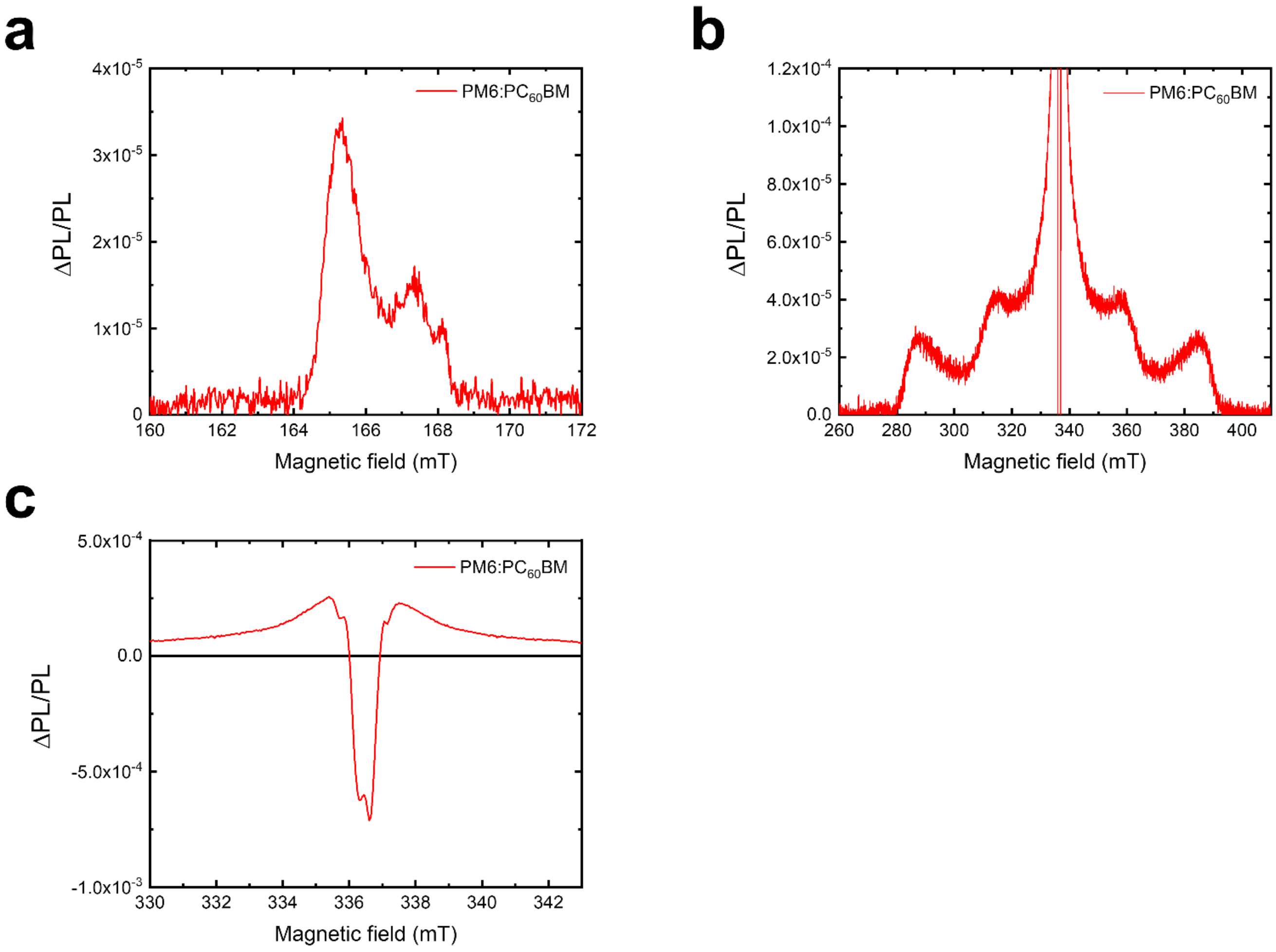

**Figure S16: (a)** The HF PLDMR spectrum of the PM6:$PC_{60}BM$ blend film. The HF PM6 $T_1$ feature appears at 165.3 mT, with a smaller shoulder at 167.2 mT. The $PC_{60}BM$ HF T1 signal is weakly visible at 168.1 mT. **(b)** The zoomed in FF PLDMR spectrum of the PM6:$PC_{60}BM$ blend film. The FF PM6 $T_1$ feature spans 280-390 mT. **(c)** The central region of the PM6:$PC_{60}BM$ blend film. Two polarons with g = 2.0012 and g = 2.0034 are visible; the former corresponds to the negative $PC_{60}BM$ polaron, whilst the latter is the PM6 positive polaron.

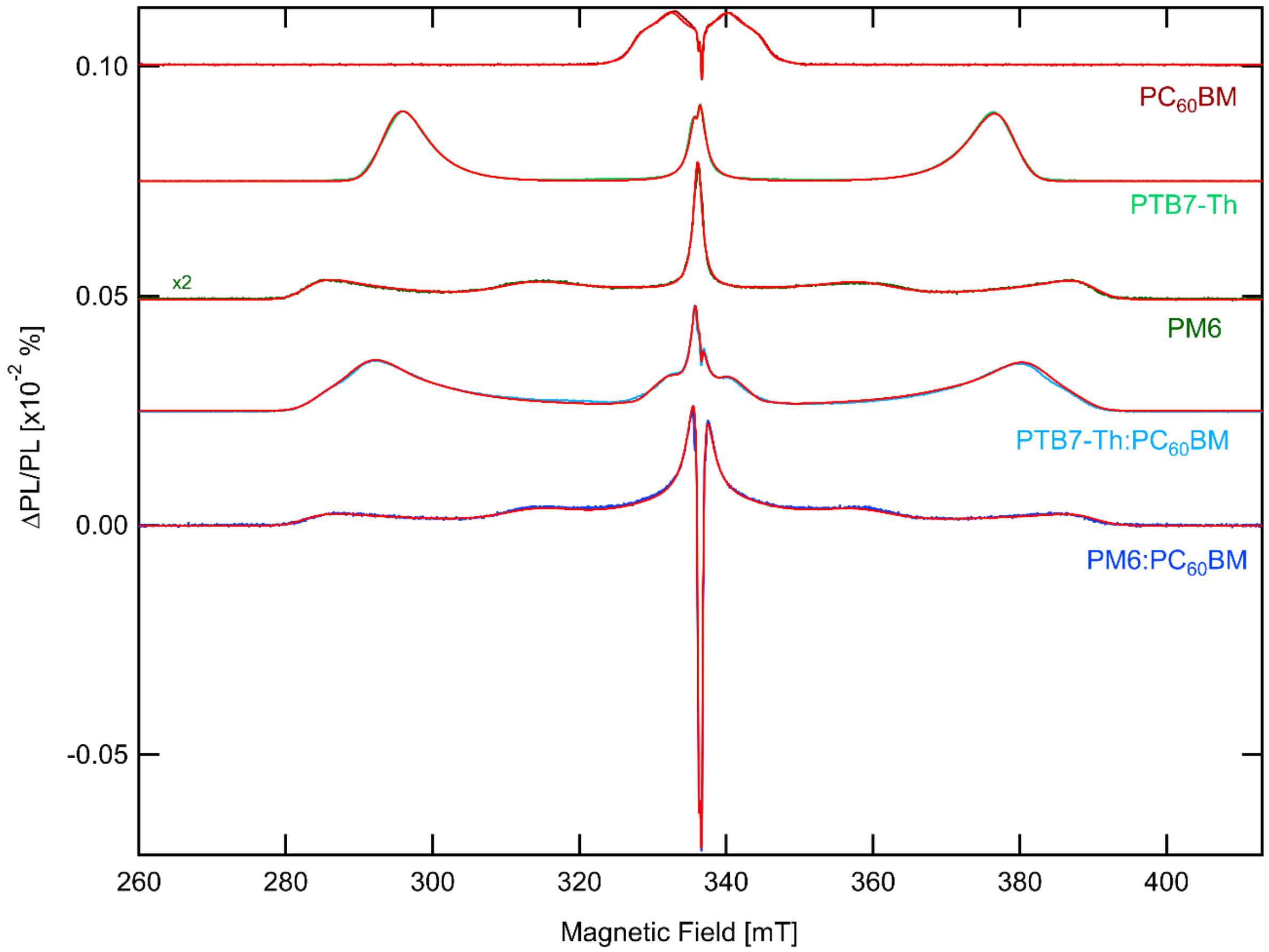

**Figure S17:** EasySpin simulations (red) of the PLDMR spectra from Figure S11. Simulation parameters are provided in Table S2.

| | **Donor** | | | | **Acceptor** | | | | **CT State** | | |
|---|---|---|---|---|---|---|---|---|---|---|---|
| **Film** | **[*D E*] /MHz** | **$\lambda_\theta$, $\lambda_\varphi$** | **LW /mT** | **Weight** | **[*D E*] /MHz** | **$\lambda_\theta$, $\lambda_\varphi$** | **LW /mT** | **Weight** | **LW/mT** | **Weight** | **g** |
| PC$_{60}$BM | | | | | [360 50] | -1, 0 | [2.5 0] | 1.00 | [0 0.3]<br>[0 0.2] | -0.03<br>-0.07 | 2.0012<br>2.004 |
| PTB7-Th | [1190 180*] | 11, 0 | [5 0] | 0.78 | | | | | [0 2]<br>[0 0.8] | 0.22<br>-0.34 | 2.0042<br>2.0047 |
| PTB7-Th:PC$_{60}$BM | [1310 180**]<br>[1470 180**] | 7.5, -4<br>3, -4 | [6 0]<br>[5 0] | 0.0037<br>0.0046 | [320 30] | -1, 0 | [2.5 0] | 0.52 | [0 1.8]<br>[0 0.35]<br>[0 0.6] | 0.48<br>-0.03<br>-0.06 | 2.0044<br>2.0012<br>2.0037 |
| PM6 | [1500 70] | 1, 2<br>5.5, 3 | [9 0]<br>[5 0] | 0.52<br>0.30 | | | | | [0 1.5] | 0.18 | 2.0045 |
| PM6:PC$_{60}$BM | [1500 70] | 0,5, 3<br>5, 3 | [9 0]<br>[5 0] | 0.42<br>0.11 | [360 50] | | [0 4] | 0.10 | [0 2.8]<br>[0 0.5]<br>[0 0.4] | 0.37<br>-0.15<br>-0.09 | 2.0025<br>2.0012<br>2.0034 |

**Table S2:** Results of EasySpin simulations with axial (*D*) and rhombic (*E*) zero-field splitting parameters, ordering factors for distribution along $\theta$ and $\phi$ ($\lambda_\theta$, $\lambda_\phi$), gaussian and lorentzian linewidh (LW), weight (according to normalized spectra) and g-factor. $PC_{60}BM$ shows two negative signals, likely corresponding to $PC_{60}BM$ anion and cation.
*: *E* value is determined from dropcast samples due to high ordering of spin-coated samples.
**: *E* value can not be determed due to high ordering, as described in the text.

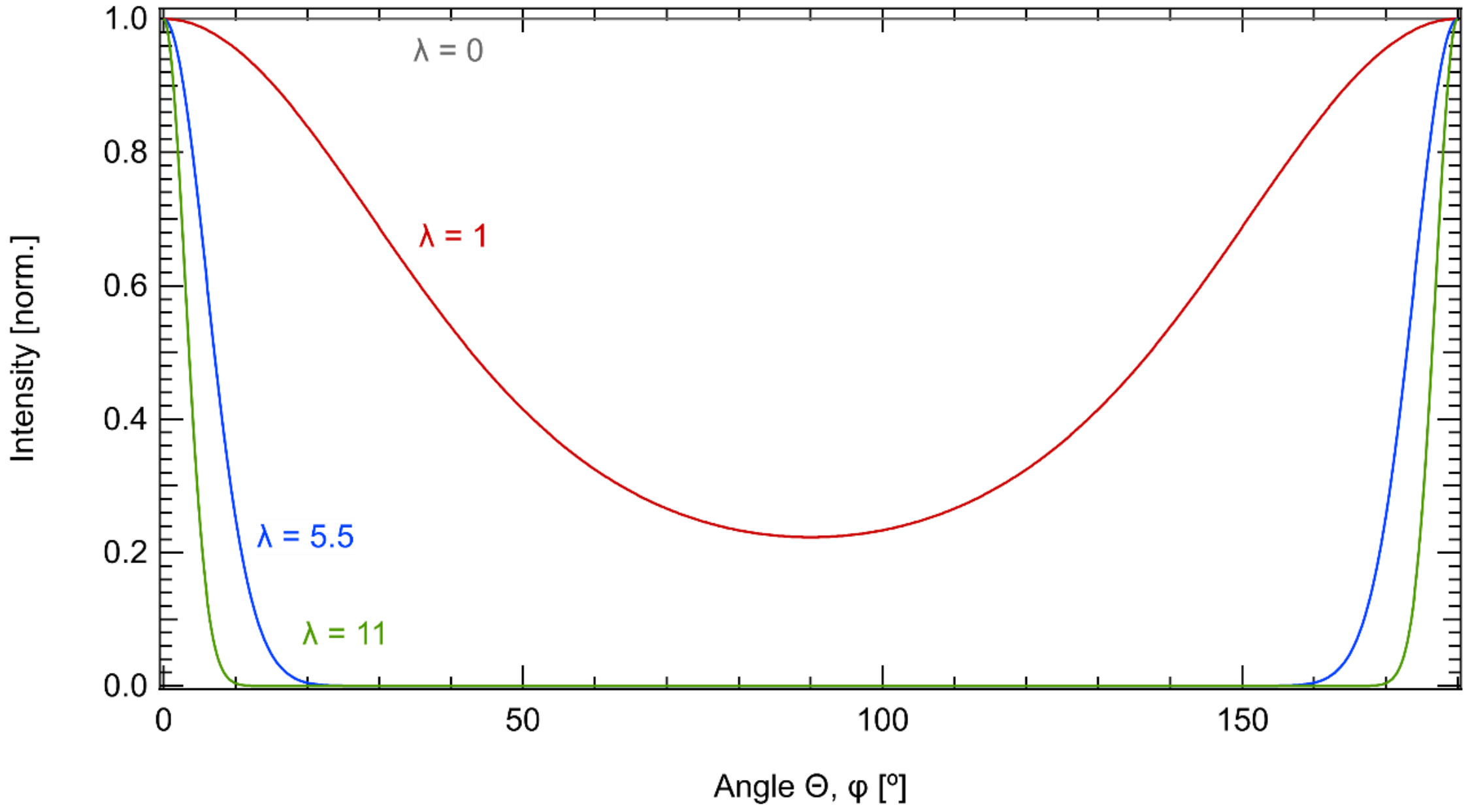

**Figure S18:** Orientational distribution of molecules in the samples for different ordering factors $\lambda$. The ordering factor $\lambda$ is given for $\theta$ and $\phi$, where $\theta$ is the angle between molecular z-axis and static magnetic field $\mathbf{B_0}$ and $\phi$ the in-plane angle. If $\lambda$ is zero, all molecular orientations occur with the same probability. A high ordering factor $\lambda_\theta$ corresponds to an extremely narrow orientational distribution of the molecules in the direction of the applied magnetic field.

**Combining TA, trEPR, and PLDMR**

Our analysis shows that through leveraging the strengths of the three techniques, a complete picture of $T_1$ generation pathways in OSCs can be obtained. For example, TA can identify $T_1$ formation on timescales spanning femtoseconds to milliseconds; this allows for the different photophysical pathways to be tracked at room temperature, thereby providing a direct comparison with real device working conditions. Though it is possible to use TA to extract more detailed information related to $T_1$ states, such as the fraction of charge carrier recombination that proceeds *via* $T_1$, this requires an advanced analysis of the correlated $T_1$ generation and charge recombination kinetics over multiple fluences[1–3]. Conversely, both magnetic resonance techniques presented here provide a more direct way to study $T_1$ states. However, the limited time resolution of these techniques necessitates the use of cryogenic temperatures to slow down the exciton and spin relaxation kinetics. In trEPR, sample excitation is provided by nanosecond laser pulses and the technique has a time resolution of hundreds of nanoseconds, whilst PLDMR is based on continuous optical excitation, thus probing steady-state populations. Therefore, whilst trEPR relies on spin polarisation to differentiate between $T_1$ populating mechanisms, the populating pathways cannot be readily resolved in PLDMR. However, the optical detection renders this method sensitive to all $T_1$ states. Furthermore, even under steady-state conditions, the spin polarisation of $T_1$ states can accumulate due to unequal recombination rates and triplet-triplet annihilation, resulting in the exceptional sensitivity of PLDMR. Therefore, the method can also detect $T_1$ states formed via non-geminate BCT. Further details on the strengths and weaknesses of each technique are reported in Table 1. Importantly, it is the combination of all three spin-sensitive techniques that enables us to fully characterise $T_1$ formation pathways in our model OSC systems.

**Discussion of PTB7-Th:$PC_{60}$BM**

In TA from the NIR region of the PTB7-Th:$PC_{60}$BM blend (Figure S4a) at 0.1-0.15 ps after photoexcitation at 700 nm, we observe two PIA features: one peaked at 1125 nm and a broader band extending towards the edge of our probe range at 1450 nm. The PIA at 1125 nm is attributed to the hole polaron located on PTB7-Th[4], whilst the band around 1450 nm is confirmed to be the residual PTB7-Th $S_1$ PIA though comparison with the TA spectrum of a neat PTB7-Th film (Figure S3). As with PM6:$PC_{60}$BM, the rapid quenching of the $S_1$ PIA indicates ultrafast electron transfer from the electron donor polymer to the fullerene. However, unlike PM6:$PC_{60}$BM, we do not observe the formation of a clear $T_1$ PIA over timescales of hundreds of picoseconds. Rather, we see an apparent broadening on the low energy edge of the hole polaron PIA. Kinetic traces from this region around 1250-1300 nm reveal a clear fluence dependence (Figure S4b), indicating a bimolecular formation mechanism. Thus, in-line with a previous report[4], we conclude that the PIA of PTB7-Th $T_1$ states, formed via the non-geminate BCT process, overlaps with the hole PIA.

We now turn to the trEPR of the PTB7-Th:$PC_{60}$BM blend film (Figure S9). At 1 μs, this blend shows a strong *ea* feature at ~346 mT, which is assigned to CT states[5]. In contrast to PM6:$PC_{60}$BM, we do not observe an evolution of the CT states into free charges by 5 μs in PTB7-Th:$PC_{60}$BM. This suggests that, at least at 80 K, the generation of free charges that are no longer magnetically interacting is slower in PTB7-Th:$PC_{60}$BM. In addition, we also observe a signal between 300-390 mT at both 1 and 5 μs; as the spectrum is significantly broader than the $PC_{60}$BM $T_1$, we attribute this feature to $T_1$ states on PTB7-Th. We verify this assignment through comparison of the ZFS parameters obtained from simulations of the neat PTB7-Th (Figure S8) and PTB7-Th:$PC_{60}$BM blend films (Table S1). The $T_1$ spectra cannot be well-described by a best-fit simulation with single *aaaeee* SOC-ISC species (Figure S8), confirming that there is more than one $T_1$ generation mechanism present. Therefore, we have simulated the $T_1$ spectra at 1 and 5 μs with two species (see Table S1); one with an *aaaeee* and the other with an *eaaeea* polarisation pattern, representing $T_1$ states formed via SOC-ISC and the geminate BCT mechanism, respectively. Thus, we are able to definitively confirm that geminate BCT $T_1$ formation is also occurring in the PTB7-Th:$PC_{60}$BM.

In the PDLMR of the neat film PTB7-Th (Figure S11, light green), we observe a $T_1$ spectrum with large ordering factor, $\lambda$. In contrast to trEPR, ordering factors are only visible in PLDMR for these materials, further discussed below. The ordering factor $\lambda$ is given for $\theta$ and $\phi$, where $\theta$ is the angle between the molecular z-axis and the applied magnetic field, and $\phi$ is the in-plane angle. If $\lambda$ is zero, all molecular orientations occur with the same probability. For PTB7-Th, the ordering for $\theta$ is $\lambda_\theta = 11$, corresponding to an extremely narrow orientational distribution of the molecules in the direction of the applied magnetic field (Figure S18). Upon blending PTB7-Th with $PC_{60}$BM (Figure S11, dark green), a broad $T_1$ feature of the polymer is clearly visible in the FF and HF signals of the blend. However, the ZFS *D*-parameter increases from $D = 1190$ MHz in pristine PTB7-Th to $D = 1310$ MHz in the blend, while the ordering factor decreases to $\lambda_\theta$ = 7.5. The change in these values suggest that blending PTB7-Th with $PC_{60}BM$ affects the polymer chain ordering in the PTB7-Th domains.

**Comparison of trEPR and PLMDR – orientation effects in triplet states**

PLDMR and trEPR are both based on the application of an external magnetic field to split the triplet sublevels and probe the spin transitions through microwave radiation. While trEPR probes all spin-polarized triplet states directly *via* detection of microwave absorption and emission, PLDMR probes only those triplet states associated with luminescence, e.g. via triplet-triplet annihilation (TTA), ground-state depletion or reverse intersystem crossing. As a result of the different detection methods, we highlighted intriguing differences from the comparison of our trEPR and PLDMR results. Specifically, PLDMR spectra showed higher ordering with respect to trEPR spectra, which show no significant preferred orientation (see e.g. PTB7-Th). We rationalise this by considering that trEPR detects all spin-polarized triplet excitons, while PLDMR is sensitive only to triplet excitons related to photoluminescence. Therefore, it seems that the PLDMR signal predominantly stems from the ordered polymer (crystalline) domains, in which TTA is efficient, rather than amorphous domains, that are also probed by trEPR. This explains why trEPR spectra show a negligible ordering factor, but also implies that the share of spin-polarized triplet excitons can even be higher in the amorphous domains, e.g., due to longer lifetimes not shortened by TTA. The same effect is visible when blending PTB7-Th with $PC_{60}BM$. In contrast to PLDMR, which shows an extreme change in the ZFS and ordering factors after blending, trEPR spectra do not show such a significant difference between neat material and blends, suggesting the scenario that the crystalline domains are mostly influenced when blending with $PC_{60}BM$. Probing particularly crystalline phases in PLDMR also explains the slight differences in ZFS parameters between trEPR and PLDMR, where PLDMR tends to show higher D values.

**Comparison trEPR and PLMDR – time evolution of triplet states**

The employed steady-state PLDMR technique does not offer relevant time resolution and, in most cases, can only be applied at very low temperatures. Pulsed PLDMR techniques with nanosecond time resolution do exist and are mainly applied to study coherent spin phenomena of luminescent color centers (spin baring defects) in wide-gap insulators or semiconductors, such as diamond, silicon carbide or hexagonal boron nitride. This is enabled by the long spin relaxation times in these materials but is not applicable to OSC blends with fast spin relaxation. The employed trEPR, however, allows investigating OSC triplet dynamics with a time resolution of hundreds of nanoseconds. From the comparison of trEPR spectra acquired at different times after the laser pulse it is possible to obtain important information about the dynamics and the spin relaxation of triplet excitons. For example, we observed small changes in the ZFS parameters of the triplets at different delays after the laser pulse, as well as a change in the spin populations of the three triplet sublevels. These observations can be explained with small variations in the molecular delocalization of the detected triplets with time (the *D* parameter is dependent on the triplet wavefunction delocalization: $D \propto r^{-3}$), and with an anisotropy in the spin relaxation times of the triplet sublevels. This anisotropy is particularly pronounced in the spectra of e.g. PM6:$PC_{60}BM$ where the polarization pattern of the ISC triplet varies from eeeaaa (1 μs) to aeaeae (5 μs) as a result of the unequal decay rates of the triplet sublevels.

## Supplementary References

1 A. J. Gillett, A. Privitera, R. Dilmurat, A. Karki, D. Qian, A. Pershin, G. Londi, W. K. Myers, J. Lee, J. Yuan, S.-J. Ko, M. K. Riede, F. Gao, G. C. Bazan, A. Rao, T.-Q. Nguyen, D. Beljonne and R. H. Friend, *Nature*, 2021, **597**, 666–671.

2 A. Rao, P. C. Y. Chow, S. Gélinas, C. W. Schlenker, C.-Z. Li, H.-L. Yip, A. K.-Y. Jen, D. S. Ginger and R. H. Friend, *Nature*, 2013, **500**, 435–439.

3 S. M. Menke, A. Sadhanala, M. Nikolka, N. A. Ran, M. K. Ravva, S. Abdel-Azeim, H. L. Stern, M. Wang, H. Sirringhaus, T.-Q. Nguyen, J.-L. Brédas, G. C. Bazan and R. H. Friend, *ACS Nano*, 2016, **10**, 10736–10744.

4 Y. Tamai, Y. Fan, V. O. Kim, K. Ziabrev, A. Rao, S. Barlow, S. R. Marder, R. H. Friend and S. M. Menke, *ACS Nano*, 2017, **11**, 12473–12481.

5 J. Niklas, S. Beaupré, M. Leclerc, T. Xu, L. Yu, A. Sperlich, V. Dyakonov and O. G. Poluektov, *J. Phys. Chem. B*, 2015, **119**, 7407–7416.